\documentclass[aps,prl,twocolumn,superscriptaddress]{revtex4-1}
\usepackage{amsmath,amsfonts,amssymb}
\usepackage{graphicx,float,calc}
\usepackage{color,bm}
\usepackage{ulem}
\usepackage{braket}
\usepackage[colorlinks,urlcolor=blue,citecolor=blue,linkcolor=blue]{hyperref}

\begin{document}

\title{Quantized Topological Anderson-Thouless Pump}

\author{Yi-Piao Wu}
\affiliation{Guangdong Provincial Key Laboratory of Quantum Engineering and Quantum Materials, School of Physics and Telecommunication Engineering, South China Normal University, Guangzhou 510006, China}

\author{Ling-Zhi Tang}
\affiliation{Guangdong Provincial Key Laboratory of Quantum Engineering and Quantum Materials, School of Physics and Telecommunication Engineering, South China Normal University, Guangzhou 510006, China}

\author{Guo-Qing Zhang}
\affiliation{Guangdong Provincial Key Laboratory of Quantum Engineering and Quantum Materials, School of Physics and Telecommunication Engineering, South China Normal University, Guangzhou 510006, China}
\affiliation{Guangdong-Hong Kong Joint Laboratory of Quantum Matter, Frontier Research Institute for Physics, South China Normal University, Guangzhou 510006, China}

\author{Dan-Wei Zhang}\thanks{danweizhang@m.scnu.edu.cn}
\affiliation{Guangdong Provincial Key Laboratory of Quantum Engineering and Quantum Materials, School of Physics and Telecommunication Engineering, South China Normal University, Guangzhou 510006, China}
\affiliation{Guangdong-Hong Kong Joint Laboratory of Quantum Matter, Frontier Research Institute for Physics, South China Normal University, Guangzhou 510006, China}

\begin{abstract}
Thouless pump with quantized transports is topologically robust against small perturbations and  disorders, while breaks down under sufficiently strong disorders. Here we propose counter-intuitive topological pumps induced by disorders in noninteracting and interacting systems. We first show an extrinsic topological pump driven by the on-site quasiperiodic potential for a two-loop sequence, where the disorder inequivalently suppresses the topology of two pump loops. Moreover, we reveal an intrinsic topological pump induced by the hopping quasiperiodic disorder from a trivial single-loop pump in the clean limit, dubbed the topological Anderson-Thouless pump (TATP) as a dynamical analogue of topological Anderson insulators. We demonstrate that the mechanism of the TATP is the disorder-induced shift of gapless critical points and the TATP can even exhibit in the dynamic disorder and interacting cases. Finally, we extend the TATP to higher-order topological systems with disorder-induced quantized corner transports. Our proposed TATPs present new members of the topological pump family and could be realized with ultracold atoms or photonic waveguides.

\end{abstract}

\date{\today}

\maketitle

{\color{blue}\textit{Introduction.---}} In 1983, Thouless proposed quantized transport of electrons in a 1D potential under slow periodic modulations in time \cite{Thouless1983}. In such Thouless pump, the quantization of pumped charge per cycle is related to the Chern number and shares the same topology as the inter quantum Hall effect in 2D \cite{Klitzing1980,Niu1984,QNiu1985,Xiao2010}, with one spatial dimension replaced by one temporal parameter. Recent studies of Thouless pump greatly enrich topological phases from fermions to wave phenomena in condensed matter and artificial systems \cite{XLQi2011,Hasan2010,DWZhang2018,Cooper2019,Goldman2016,Ozawa2019}. Topological pumps have been realized using ultracold atoms \cite{Lohse2015,Nakajima2016,HILu2016,Schweizer2016,Nakajima2021,Fabre2022,Minguzzi2022,LWang2013a}, photonic waveguides \cite{Kraus2012a,Cerjan2020,QCheng2022,WLiu2022,YKe2016}, acoustic and mechanical metamaterials \cite{Cheng2020,Grinberg2020}. The concept of Thouless pump has been extended to spin pump \cite{LFu2006,Schweizer2016}, interacting topological pumps \cite{TSZeng2015,Taddia2017,Li2017,GonzalezCuadra2019,Kuno2020,Marks2021,Walter2022,Gawatz2022,YQian2011,YKe2017}, 2D topological pump related to the 4D quantum Hall effect \cite{ZhangSC2001,Kraus2013b,Lohse2018,Zilberberg2018}, Floquet-Thouless energy pump \cite{Kolodrubetz2018}, non-linear \cite{Juergensen2021,Juergensen2022,Juergensen2022b,Mostaan2021,QFu2022} and non-Abelian \cite{Brosco2021,OYou2022,YSun2022} Thouless pumps, and higher-order topological
pump \cite{Wienand2022,Benalcazar2022b,BXie2021}.

An essential property of the Thouless pump and other topological effects is their stability against disorders as long as the energy gap for adiabaticity remains intact \cite{XLQi2011,Hasan2010,DWZhang2018,Cooper2019,Goldman2016,Ozawa2019}. However, disorder usually acts as a suppressor due to the Anderson localization \cite{Anderson1958} and breaks down topological pumps when the disorder-driven gap closing happens \cite{Qin2016,Wauters2019,Hayward2021}. The robustness and breakdown of Thouless pump under on-site random and quasiperiodic disorders have been experimentally studied \cite{Cerjan2020,Nakajima2021}. Remarkably, a disorder-induced non-quantized pump for a specific sequence connecting two pump loops was demonstrated with noninteracting atoms in a quasiperiodic optical lattice (OL) \cite{Nakajima2021}. A striking disorder-induced topological phase known as the topological Anderson insulator (TAI) was revealed \cite{JLi2009}, which have been theoretically extended \cite{Groth2009,HJiang2009,CZChen2015,HMGuo2010,Altland2014,Mondragon-Shem2014,Titum2015,Sriluckshmy2018,RChen2019,DWZhang2020,XWLuo2019,QLin2022,WHong2020,XSWang2020,GQZhang2021,KLi2021,CLi2020,YBYang2021,WZhang2021} and experimentally observed with ultracold atoms and photonic crystals \cite{Meier2018a,Stutzer2018,GGLiu2020}. In the existing works, an unexplored question is whether a intrinsic topological pump can be induced by disorders and exhibit in different systems. This important question involves the interplay among topology, disorder and dynamics in condensed matter and artificial systems.

In this Letter, we explore disorder-induced quantized topological pumps in noninteracting and interacting systems with different disorders. We first show an extrinsic topological pump driven by the on-site quasiperiodic potential for a specific two-loop pumping sequence \cite{Nakajima2021}, where the disorder suppresses the topology of two pump loops inequivalently. Moreover, we reveal a quantized topological pump intrinsically induced by the hopping quasiperiodic disorder from a trivial single-loop pump in the clean regime, which is dubbed topological Anderson-Thouless pump (TATP) and can be viewed as a dynamical analogue of TAIs. The mechanism of the TATP is found to be the disorder-induced shift of gapless critical points. We further demonstrate the exhibition of TATP in the dynamic disorder and interacting cases, and propose the higher-order TATP with disorder-induced quantized corner transports in 2D systems. Our proposed TATPs can already be realized with ultracold atoms or photonic waveguides, which present new members of the topological pump family.

\begin{figure*}[t]
\centering
\includegraphics[width=0.9\textwidth]{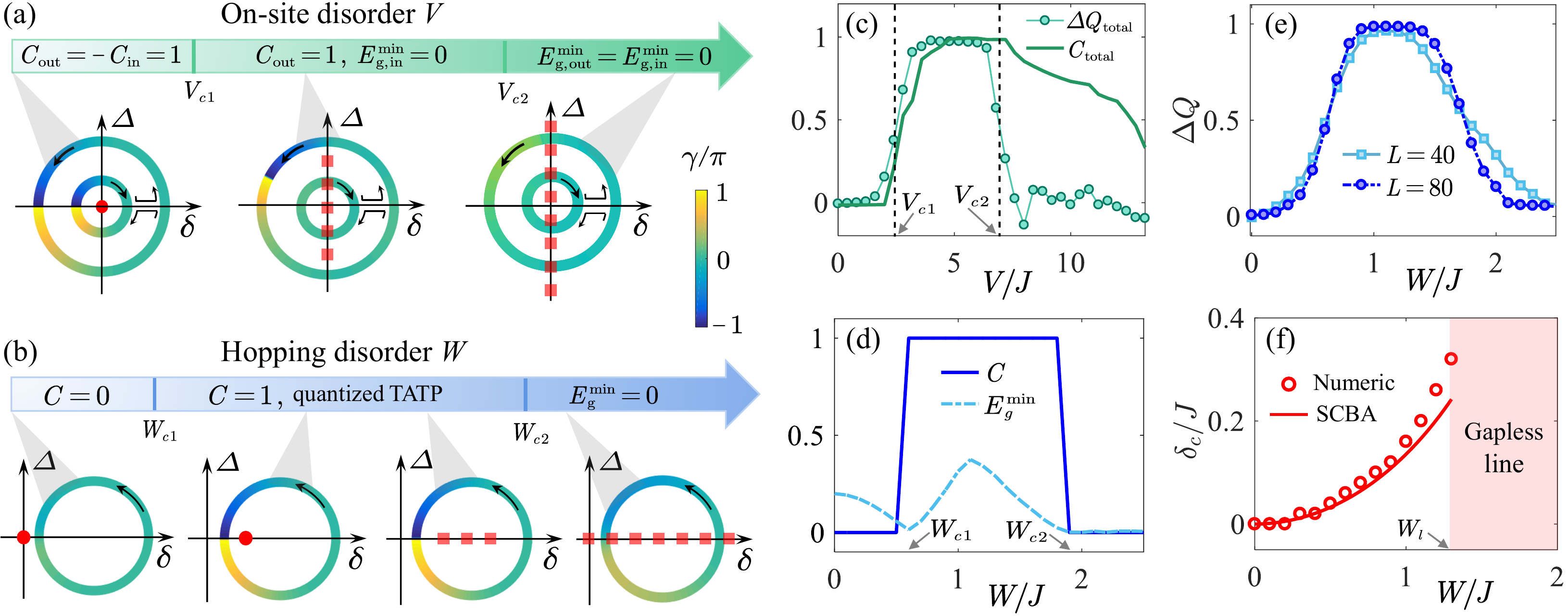}
\caption{(Color online) Schematic of extrinsic and intrinsic topological pumps induced by on-site (a) and hopping (b) quasiperiodic disorders, respectively. Pump cycles are sketched in the $\delta$-$\Delta$ plane, with colors and directions denoted by the Berry phase $\gamma$ and the arrows, red points and dashed lines for gapless regimes. Critical disorder strengths are $\{V_{c1},V_{c2}\}\approx\{2.5,6.8\}$, and $\{W_{c1},W_{c2}\}\approx\{0.6,1.9\}$. (c) Total Chern number $C_{\text{total}}$ and pumped charge $\Delta Q_{\text{total}}$ for $\{L=80,T=400\}$ as functions of $V$. Dashed lines denote critical strengths, after which the inner and outer loops across gapless lines as shown in (a), respectively. (d) Chern number $C$ and minimum energy gap $E_g^{\text{min}}$ as a function of $W$. (e) $\Delta Q$ as a function of $W$ for $\{L=40,T=200\}$ and $\{L=80,T=400\}$. (f) Location of gapless point $\delta_c$ (numeric and SCBA approach) as a function of $W$ for $W<W_l\approx1.3$. When $W>W_l$, the gapless regime becomes a line. Other parameters are $J=1$, $\beta=0$, $U=0$, and $\{R_{\Delta},R_{\delta},\delta_s\}=\{5,5,0\}$, $\{0.7,0.7,0\}$, $\{1,0.45,0.5\}$ for outer, inner, and single loops in (a,b), respectively. Loops from left to right are $V=\{0,5,12\}$ in (a), and $W=\{0,1.2,1.6,2.0\}$ in (b). The lattice size is $L=40$ for computing $\gamma$ and $C$, and $L=800$ for $E_g^{\text{min}}$.}
\label{fig1}
\end{figure*}

{\color{blue}\textit{Model and topology.---}}Let us consider spinless fermions in a 1D optical superlattice with a unit cell of two sites \cite{Lohse2015,Nakajima2016,HILu2016,Schweizer2016,Nakajima2021,Fabre2022,Minguzzi2022}. Atoms at the site $j$, described by creation and annihilation operators $\hat{c}^{\dagger}_j$ and $\hat{c}_j$, move along the lattice with staggered potential $\Delta$, alternated hopping, and Hubbard interaction $U$. The tight-binding Hamiltonian reads $\hat{H}=\hat{H}_0+ U\sum_j\hat n_j\hat n_{j+1}$ with
\begin{equation}\label{Ham1}
\hat{H}_0 =-\sum_{j} (J_j \hat{c}^{\dagger}_j \hat{c}_{j+1} + \text{H.c.})
+ \sum_j \left[(-1)^j\Delta+V_j\right] \hat{n}_j.
\end{equation}
%
Here $\hat n_j = \hat c^{\dagger}_j \hat c_j$ is the particle number operator, and the hopping strengths $J_{j\in \text{even}}=J+\delta+W_j$ and $J_{j\in \text{odd}}=J-\delta$ with parameters $\delta$ and $J\equiv1$ ($\hbar/J\equiv1$) as the energy (time) unit. We first consider the static on-site and intra-cell hopping disorders of the quasiperiodic forms $V_j=V\cos(2\pi\alpha j+\beta)$ and $W_j=W\cos(2\pi\alpha j+\beta)$ with strengths $V$ and $W$, respectively. Here $\alpha=(\sqrt{5}-1)/2$ denotes the incommensurate modulation and $\beta$ is a shift phase. The hopping and on-site disorders preserve and break the (sublattice) chiral symmetry when $\Delta=0$, respectively, both of which have been realized for ultracold atoms \cite{Meier2018a,Nakajima2021}.    We focus on the noninteracting pumping in the lattice size $L$ and particle number $N=L/2$ at half filling, and consider the interacting case later.

A pumping cycle on the $\Delta$-$\delta$ plane can be parametrized by $\phi=\phi(t)=2\pi t/T\in[0,2\pi]$ over time $t$, with period $T$ and dynamical parameters $\Delta(t)=R_{\Delta}\sin\phi(t)$ and $\delta(t)=R_{\delta}\cos\phi(t)+\delta_s$. The slow change of $\phi(t)$ induces a Thouless pump with the pumped charge after a cycle
\begin{eqnarray}
\Delta Q= Q(T)-Q(0)=\int_0^{T} dt \langle\Psi(t)|\mathcal{\hat{J}}(t)|\Psi(t)\rangle,
\end{eqnarray}
where $\mathcal{\hat{J}}=\frac{i}{L}\sum_{j=1}^{L} J_j \hat{c}^{\dagger}_j \hat{c}_{j+1} + \text{H.c.}$ is the averaged current operator. In the presence of disorder or interaction \cite{Niu1984,QNiu1985,Xiao2010}, one can impose the twisted periodic boundary condition (PBC) by introducing a twist phase $\theta\in[0,2\pi]$ to the hopping term $\hat{c}^{\dagger}_j \hat{c}_{j+1}\rightarrow e^{i\theta/L}\hat{c}^{\dagger}_j \hat{c}_{j+1}$ in the Hamiltonian in Eq. (\ref{Ham1}), denoted as $\hat{H}_{\theta}$. The topology of an adiabatic pump can be defined by the ground state $\vert \Psi_g\rangle$ of Hamiltonian $\hat{H}_{\theta}$ on the torus spanned by $(\theta,\phi)$, or equivalently $(\theta,t)$. The Berry phase reads
\begin{eqnarray}
\gamma=\int_0^{2\pi} d \theta\bra{\Psi_g}i \partial_{\theta}\Psi_g\rangle= \sum_{n=1}^{N} \int_0^{2\pi} d\theta \bra{\psi_n} i\partial_{\theta} \psi_n\rangle,
\end{eqnarray}
where $\vert \Psi_g\rangle$ reduces to single-particle eigenstates $\ket{\psi_n}$ populated up to half-filling in the free-fermion case. The Chern number $C$, as the winding number of the Berry phase, characterizes the topology of a Thouless pump \cite{Thouless1983}
\begin{eqnarray}
C =\int_0^{2\pi} \frac{d\phi}{2\pi}\partial_{\phi} \gamma=\int_0^{T} \frac{dt}{2\pi}\partial_{t}\gamma.
\end{eqnarray}
In the thermodynamic and adiabatic limits, the current $\langle\mathcal{\hat{J}}(t)\rangle=\int_{0}^{2\pi}\frac{d\theta}{2\pi}\langle\mathcal{\hat{J}}_\theta(t)\rangle =\int_{0}^{2\pi}\frac{d\theta}{2\pi}\bra{\Psi_g(t)}\partial_{\theta}\hat{H}_{\theta}\ket{\Psi_g(t)}=\frac{1}{2\pi}\partial_t\gamma$ \cite{Niu1984}. Thus the quantization of pumped charge is guaranteed by the Chern number $\Delta Q=C$ in an ideal Thouless pump \cite{Thouless1983}. This pumped charge corresponds to the center-of-mass shift of an atomic gas \cite{Lohse2015,Nakajima2016,HILu2016,Schweizer2016,Nakajima2021,Fabre2022,Minguzzi2022,LWang2013a}, which is quantized in units of the lattice spacing.

{\color{blue}\textit{Disorder-induced topological pumps.---}} In the absence of disorder, the free-fermion Hamiltonian $\hat{H}_0$ in Eq. (\ref{Ham1}) describes the Rice-Mele model \cite{Rice1982} with a gapless point located at $\Delta=\delta=0$, which is also a singularity of Berry phase on the $\Delta$-$\delta$ plane. A quantized topological pump requires the parameter cycle $\phi(t)$ adiabatically encircling the gapless point while keeping energy gap open. Namely, the minimum energy gap during a pump cycle is $E_g^{\text{min}}=\min_{\phi}[E_{N+1}(\phi)-E_{N}(\phi)]>0$. The topological pump is robust but will break down owing to the disorder-driven preclusion of adiabaticity \cite{Wauters2019,Hayward2021,Note1}. In the following, we propose extrinsic and intrinsic topological pumps induced by on-site and hopping disorders, respectively.

We first consider a specific pump sequence connecting two loops of opposite directions under on-site disorders \cite{Nakajima2021}, as depicted in Fig. \ref{fig1}(a). In the clean and weak disorder regimes with $0\leqslant V\leqslant V_{c1}$, the energy gap closes only at $(\Delta_c,\delta_c)=(0,0)$ and the Chern numbers for outer and inner loops are $C_{\text{out}}=-C_{\text{in}}=1$. There is no net pumped charge as the total Chern number $C_{\text{total}}=C_{\text{out}}+C_{\text{in}}=0$. When $V>V_{c1}$, the gapless regime becomes a line along the $\Delta$ axis and extends when increasing $V$. For two proper loops that are far apart \cite{Note1,Nakajima2021} and under moderate disorder $V_{c1}<V<V_{c2}$, the pump along the inner loop becomes trivial as its minimum gap $E_{\text{g,in}}^{\text{min}}=0$, while the outer loop remains gapped and nontrivial with $C_{\text{out}}=1$. In this case, a topological pump with $C_{\text{total}}=1$ can be extrinsically induced by the on-site disorder, as shown in Figs. \ref{fig1}(a,c). For strong disorder $V>V_{c2}$, the gapless line across the outer loop and then $E_{\text{g,out}}^{\text{min}}=0$, such that the total pump sequence becomes trivial with non-quantized pumped charge. We perform the full-time simulation of the two-loop pumping and compute total pumped charge $\Delta Q_{\text{total}}$ for typical systems \cite{Nakajima2021} in Fig. \ref{fig1}(c), which shows the nearly quantized charge pumping for moderate $V$.

\begin{figure}[t]
\centering
\includegraphics[width=0.45\textwidth]{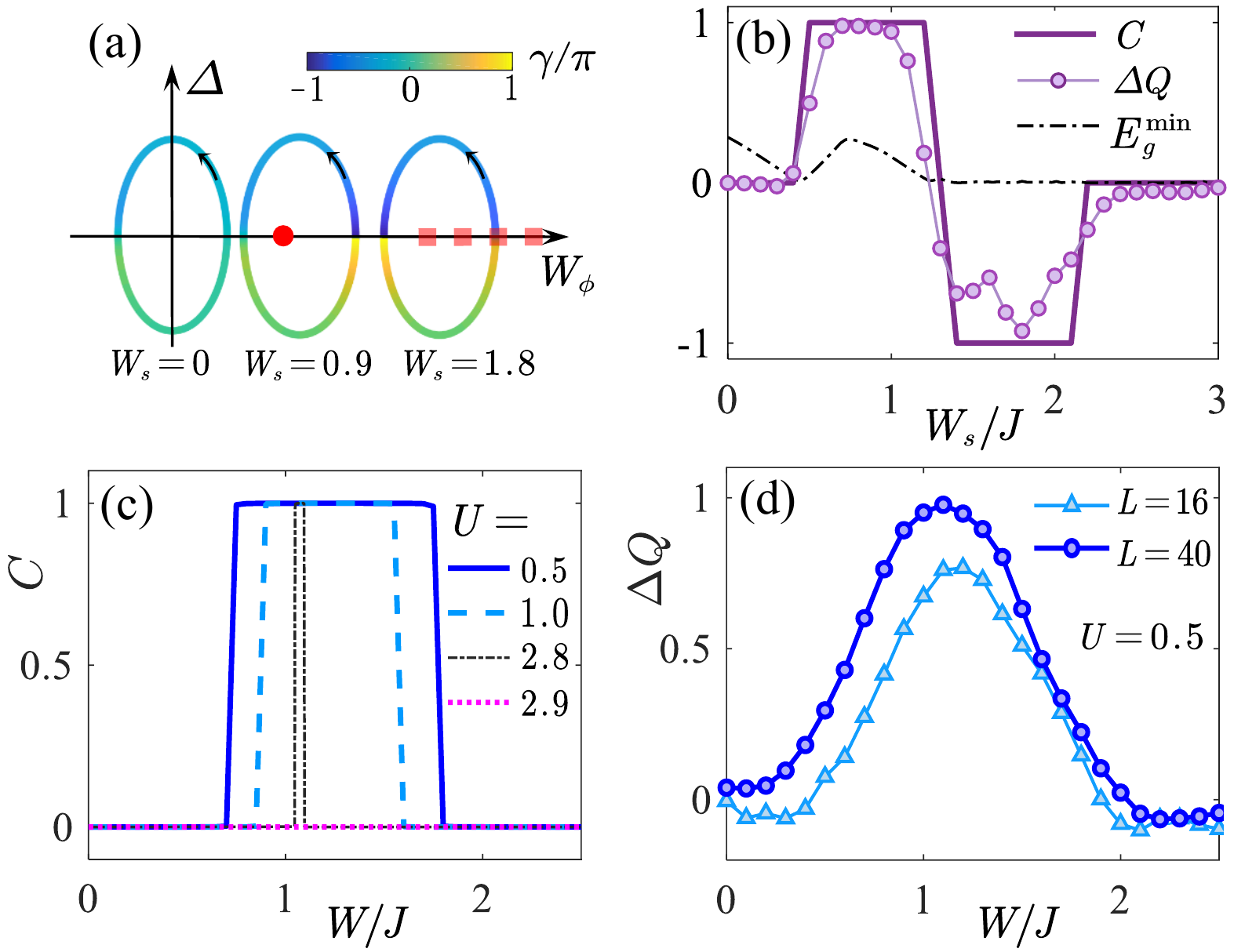}
\caption{(Color online) (a) Schematic of TATP in the $W_{\phi}$-$\Delta$ plane with three pump cycles colored by the Berry phase $\gamma$ for fixed $\{R_{\Delta},R_W,\delta\}=\{1,0.4,0.1\}$ and various $W_s$. Red point and dashed line denote the gapless regime. (b) $C$, $E_g^{\text{min}}$ and $\Delta Q$ (for $\{L=80,T=400\}$) as a function of disorder strength $W_s$ for $U=0$. (c) $C$ as a function of $W$ for varying interaction strengths $U=0.5,1.0,2.8,2.9$. (d) $\Delta Q$ as a function of $W$ for $\{L=16,T=100\}$  and $\{L=40,T=200\}$.
}
\label{fig2}
\end{figure}

We now propose an intrinsic topological pump induced by hopping disorders from a trivial single-loop pump. As shown in Fig. \ref{fig1}(b), we choose such a trivial pump loop with $C=0$ in the clean limit, without encircling the gapless point at the origin of the $\Delta$-$\delta$ plane. By increasing disorder strength $W$, we find the gapless point moves along the $\delta$ axis to $\delta_c>0$ and enters the cycle when $W_{c1}<W<W_l$, giving rise to a topological pump with $C=1$. The gapless regime becomes a line when $W>W_l$. However, the topological and gapped nature of the pump preserves as the gapless line does not cross the loop. The Chern number $C$ and minimum gap $E_g^{\text{min}}$ as a function of $W$ are shown in Fig. \ref{fig1}(d). The results indicate the topological pump induced by moderate hopping disorder. We simulate the time evolution of the single-loop pump and compute the pumped charge $\Delta Q$ for typical systems of size $L=\{40,80\}$ and period $T=\{200,400\}$ \cite{Lohse2015,Nakajima2016,HILu2016,Schweizer2016,Nakajima2021}. As shown in Fig. \ref{fig1}(e), we obtain $\Delta Q\approx0$ when $W=0$ and $\Delta Q\approx1$ when $W_l<W<W_{c2}$, which is quantized in the $L\rightarrow\infty$ limit \cite{Note1}. Thus, we reveal a quantized topological pump driven by quasiperiodic disorder from a trivial pump in the clean regime. Such a disorder-induced intrinsic topological pump is dubbed TATP and can be viewed as a dynamical version of TAIs \cite{JLi2009}. During the time evolution of the TATP, most of adiabatic eigenstates are localized \cite{Note1}. For $W>W_{c2}$, the TATP breaks down as the gapless line crossing the loop with $E_g^{\text{min}}=0$. The breakdown of the TATP could be due to the moving of the gapless point outside of a small pump loop \cite{Note1}.

The emergence of TATP is due to the disorder-induced shift of the gapless point.
To analyze this mechanism, we use the self-consistent Born approximation (SCBA) \cite{Groth2009,HJiang2009,CZChen2015}. The weak and moderate disorders can be accounted as the self-energy term $\Sigma(W)$ to renormalize the clean Hamiltonian, which satisfies the self-consistent equation
\begin{equation} \label{SCBA-Eq}
\frac{1}{E_f-\mathcal{H}_0(k)-\Sigma(W)}=\langle\frac{1}{E_f-\mathcal{H}_{\text{eff}}(k,W)}\rangle_q,
\end{equation}
where $E_f=0$ is the Fermi energy, $\mathcal{H}_0(k)=[J+\delta+(J-\delta)\cos(k)]\sigma_x+(J-\delta)\sin(k) \sigma_y + \Delta \sigma_z$ is the Bloch Hamiltonian in the clean limit with the Pauli matrices $\sigma_{x,y,z}$, the self-energy is simplified as $\Sigma(W)=\Sigma_x(W)\sigma_x$ due to the symmetry of the Hamiltonian, $\mathcal{H}_{\text{eff}}=\mathcal{H}_0(k)+W\cos(2\pi\alpha q)\sigma_x$ denotes the effective Hamiltonian renormalized by the quasiperiodic disorder with the index $q=1,2,...,N_q$, and $\braket{\cdots}_q$ denotes averaging over $N_q$ samples. By numerically solving Eq. (\ref{SCBA-Eq}) for given $W$, we obtain the renormalized hopping modulation $\tilde{\delta}=\delta+\Sigma_x(W)$ and the location of gapless point at $\delta_c=\tilde{\delta}=0$, as plotted in Fig. \ref{fig1}(f). The SCBA analysis agrees well with the numerical result of the gapless point.

{\color{blue}\textit{Dynamic disorder and interacting cases.---}} The TATP can even exhibit in the dynamic disorder and interacting cases. We first consider a more complex pump loop achieved by the time-varying hopping disorder strength $W_{\phi}=W_s+R_W\cos\phi(t)$ and $\Delta=R_{\Delta}\sin\phi(t)$, while $\delta$ is fixed. Three typical pump loops for different overall disorder strengths $W_s$ and the gapless regime on the $W_{\phi}$-$\Delta$ plane are depicted in Fig. \ref{fig2}(a). In this dynamic disorder case, the gapless point and line are separatively located on the $W_{\phi}$ axis, and their locations (regimes) are fixed for given loop parameters $(R_{\Delta},R_{W})$. Consider a trivial pump loop with $C=0$ for $W_s=0$. By increasing $W_s$, the pump loop moves along the $W_{\phi}$ axis and encircles the gapless point. This gives rise to the TATP with $C=1\approx\Delta Q$ under moderate disorder $0.5\lesssim W_s\lesssim1.3$, as shown in Fig. \ref{fig2}(b). The TATP breaks down when the loop crosses the gapless line (or moves apart from the gapless point). Note that the pump loop with $C=-1$ and $E_g^{\text{min}}=0$ when $1.3\lesssim W_s\lesssim2.2$ takes non-quantized pumped charge due to the preclusion of adiabaticity.

\begin{figure}[t]
\centering
\includegraphics[width=0.45\textwidth]{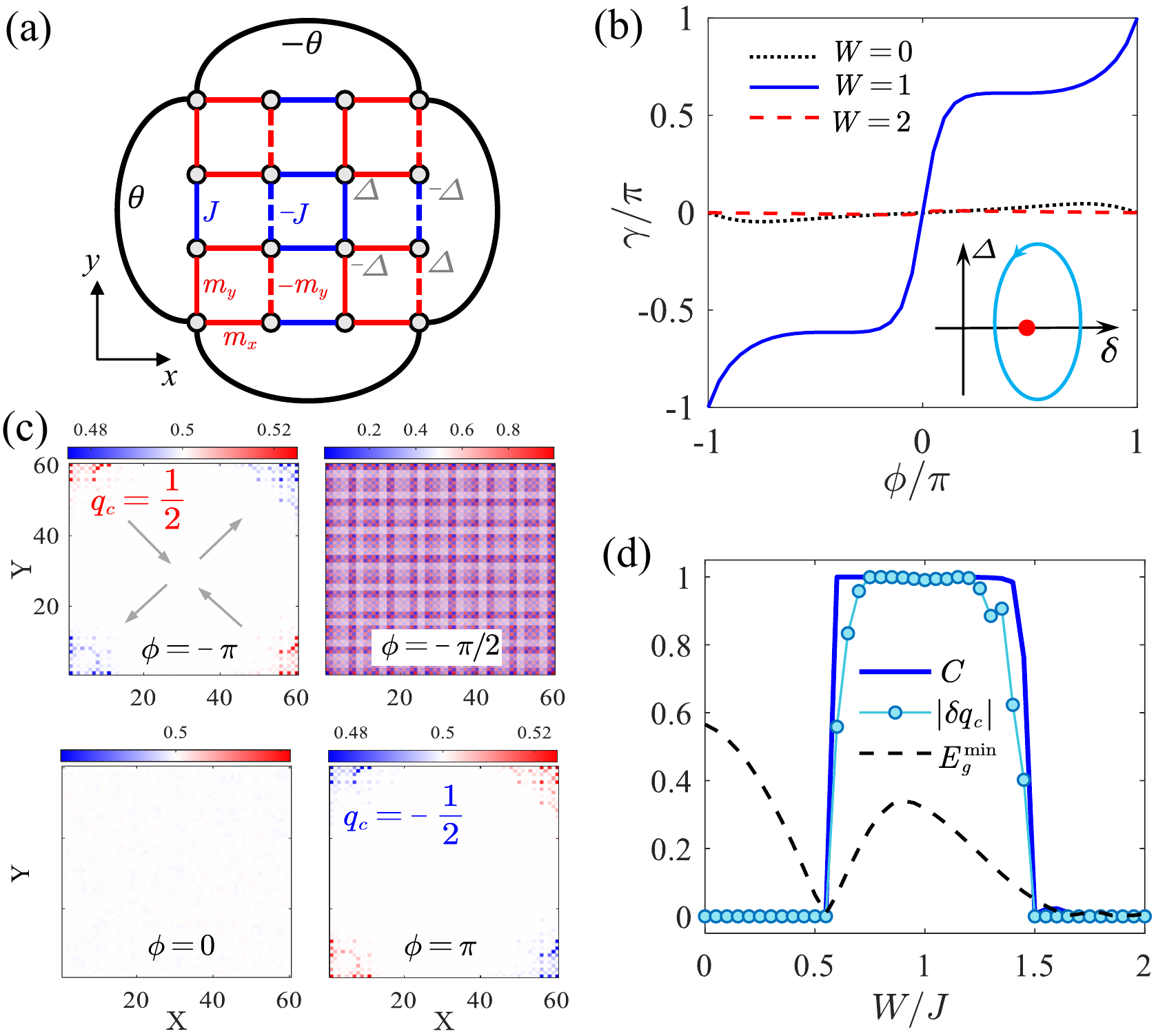}
\caption{(Color online) (a) Schematic of the higher-order topological model on a square lattice with staggered hopping and potential under the corner PBC by connecting corner sites with phase $\theta$. (b) $\gamma$ as a function of pump parameter $\phi$ during a cycle for $W=0,1,2$. The inset shows the pump loop for $W=1$. (c) Density evolution with quantized corner-to-corner charge transport under OBC for $W=1$ and $L=60$ during a pump cycle. (d) $|\delta q_c|$ (OBC), $C$ and $E_g^{\text{min}}$ (PBC) as a function of $W$. Other parameters are $\{R_{\Delta},R_{\delta},\delta_s\}=\{1,0.4,0.6\}$.}
\label{fig3}
\end{figure}

The single-loop TATP in Fig. \ref{fig1}(b) can exhibit in the presence of interactions. In the interacting case, the Chern number $C$ are obtained in terms of many-body ground states. Using numerical exact diagonalization for the lattice of $L=16$ and static disorder strength $W=1.2$, we obtain $C$ as a function of the interaction strength $U$ in Fig. \ref{fig2}(c). The disorder regime for the topological pump shrinks as increasing $U$. However, the TATP with $C=1$ preserves when $U<U_c$ with $U_c\approx2.9$. Using the density matrix renormalization group and its time-dependent methods \cite{White1992,Schollwoeck2011,Note1}, we simulate the pumping evolution up to $L=40$ and compute pumped charge $\Delta Q$ as a function of $W$ in Fig. \ref{fig2}(d), which shows the interacting TATP with $\Delta Q\approx1$ induced by moderate disorder.

{\color{blue}\textit{Higher-order TATP in 2D systems.---}} We can extend the TATP to higher-order topological systems \cite{Benalcazar2017a,Benalcazar2017b,FLiu2017,CLi2020,YBYang2021,WZhang2021,Coutant2020,Benalcazar2022}. We consider the hallmark model of the higher-order topological phase in a 2D superlattice \cite{Benalcazar2017a,Benalcazar2017b,FLiu2017}, as depicted in Fig. \ref{fig3}(a). It consists of staggered hopping amplitudes $J_{\eta}= \{J,m_{\eta}\}$ with $J\equiv1$ in two spatial dimensions $\eta=x,y$ ($x,y=1,2,...,L$), and a Peierls phase $\pi$ along vertical bonds. One can add a staggered potential $\pm\Delta$ in a cross-diagonal arrangement to realize the high-order Thouless pumping \cite{Wienand2022}. The total Hamiltonian reads
\begin{eqnarray}\label{Ham2}
\hat{H}_{2D}=~&&-\sum_{x,y} [J_x \hat{c}^{\dagger}_{x,y} \hat{c}_{x+1,y}+(-1)^{x}J_y \hat{c}^{\dagger}_{x,y} \hat{c}_{x,y+1} +\text{H.c.}] \nonumber  \\
&&+\Delta\sum_{x,y} (-1)^{x+y} \hat{n}_{x,y}.
\end{eqnarray}
For concreteness, we consider the hopping with quasiperiodic disorders $m_{\eta}=J+\delta+W\cos(2\pi\alpha \eta)$.

A pump cycle in this 2D system can be achieved via the adiabatic parameter $\phi(t)$ from $\delta(t)=R_{\delta}\cos\phi(t)+\delta_s$ and $\Delta(t)=R_{\Delta}\sin\phi(t)$ [see the inset of Fig. \ref{fig3}(b)]. Following Ref. \cite{Wienand2022}, we impose the corner PBC to characterize the topology of the higher-order pump under disorders. As illustrated in Fig. \ref{fig3}(a), a twist phase $\theta$ onto a pair of four corner-connecting links is added onto the model, denoted by the Hamiltonian $\hat{H}_{c}(\theta)$. This gives the total Hamiltonian $\hat{H}_{\text{tot}}(\theta)=\hat{H}_{2D}^{\text{OBC}}+\hat{H}_{c}(\theta)$ with the Berry phase $\gamma$ and Chern number $C$ in terms of the ground state wave function of $\hat{H}_{\text{tot}}(\theta)$ \cite{Wienand2022,Note1}. We compute $\gamma$ as a function of $\phi$ for disorder strengths $W=0,1,2$ and fixed $\{R_{\Delta},R_{\delta},\delta_s\}$ in Fig. \ref{fig3}(b). In the clean (strong disorder) regime, the winding of $\gamma$ during a trivial pump cycle is vanishing and thus $C=0$. In moderate disorder regime ($W=1$), however, the pump loop becomes nontrivial with $C=1$ as the winding of $\gamma$ is $2\pi$. The Chern number characterizes the higher-order topological pumps \cite{Wienand2022,Benalcazar2022b,BXie2021}. In Fig. \ref{fig3}(c), the density evolution of the pump under the OBC and at half filling for $W=1$ shows the transport of four corner-localized fractional charges $q_c$ \cite{Note1}, two with $q_c=\frac{1}{2}$ and two with $q_c=-\frac{1}{2}$. This yields a quantized total change of the corner charge $|\delta q_c|=C=1$. We plot $|\delta q_c|$, $C$ and $E_g^{\text{min}}$ as a function of $W$ in Fig. \ref{fig3}(d). The results indicate the higher-order TATP with quantized corner-to-corner charge transport in the 2D system.

{\color{blue}\textit{Discussion and conclusion.---}} We make some remarks before concluding. i) We also study the random disorder effect on the single-loop pump \cite{Note1}, but do not find the TATP owing to the quick preclusion of adiabaticity induced by the random disorder. ii) Although we consider the configuration with $\beta=0$, the TATP is independent of the choice of $\beta$ \cite{Note1}. iii) The finite-size scalings of Chern number, minimum energy gap and pumped charge are used to demonstrate quantized TATPs in both noninteracting and interacting cases \cite{Note1}. iv) With numerical simulations \cite{Note1}, we show the noninteracting TATP can also be observed with light in waveguide arrays \cite{Kraus2012a,Cerjan2020,QCheng2022,WLiu2022,YKe2016}.

In summary, we have uncovered an intrinsic disorder-induced topological pump, dubbed the TATP as a dynamical version of TAIs. We have demonstrated that the TATP is generic from static to dynamical disorder cases, from noninteracting to interacting regimes, and even exhibit in higher-order topological systems. The proposed TATPs could be realized using ultracold atoms or photonic waveguides and extended to other systems in future works, such as $Z_2$ spin pump \cite{Schweizer2016}, non-linear \cite{Juergensen2021,Juergensen2022,Juergensen2022b,Mostaan2021,QFu2022} and non-Abelian \cite{Brosco2021,OYou2022,YSun2022} Thouless pumps.

\acknowledgments
This work was supported by the National Natural Science
Foundation of China (Grants No. 12174126 and No. 12104166), the Key-Area Research and Development Program of Guangdong Province (Grant No. 2019B030330001), and the Guangdong Basic and Applied Basic Research Foundation (Grants No. 2021A1515010315 and No. 2020A1515110290).

Y.P.W. and L.Z.T. contributed equally to this work.

\bibliography{reference}

\clearpage


\section{Supplemental Materials}

\subsection{A. Robustness of topological pumping under quasiperiodic disorders}

\begin{figure}[!h]
\centering
\includegraphics[width=0.46\textwidth]{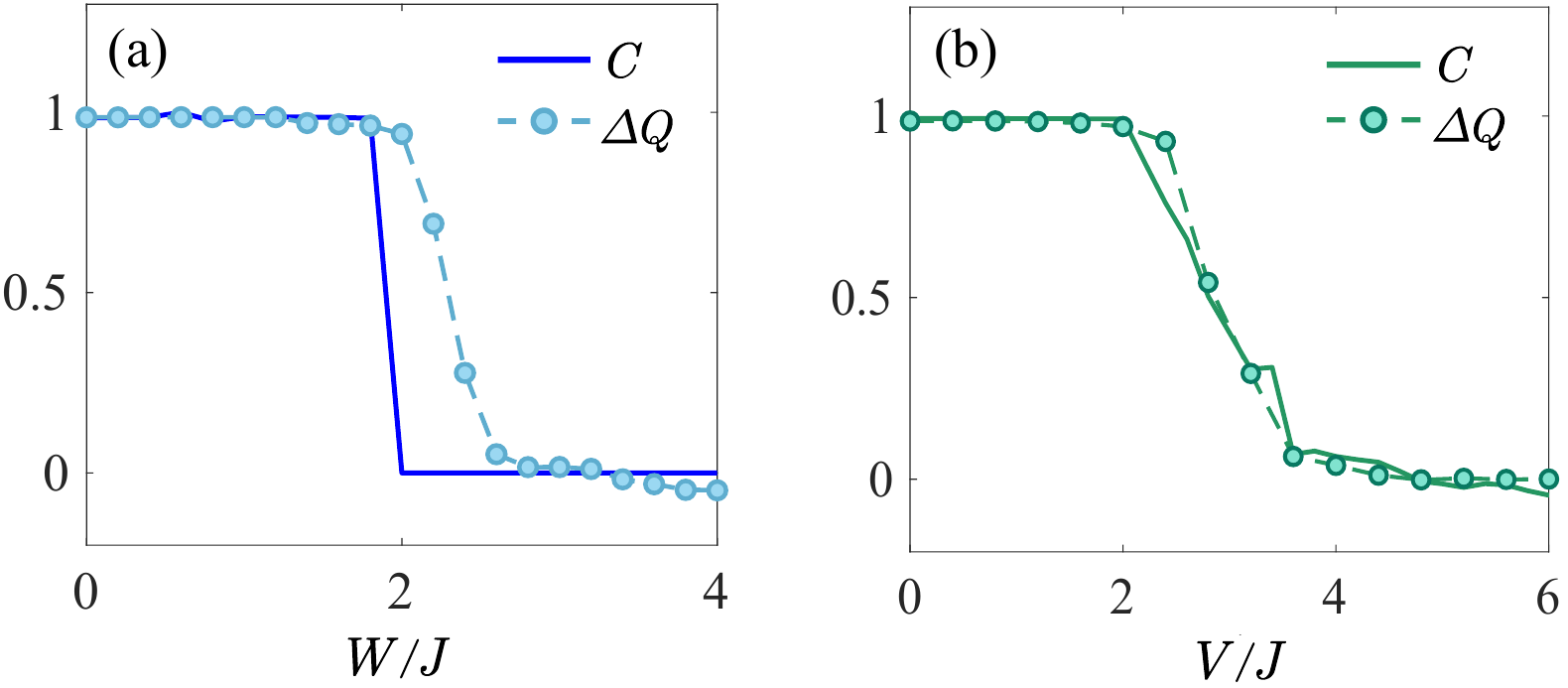}
\caption{(Color online) Chern number $C$ and pumped charge $\Delta Q$ as functions of (a) the hopping disorder strength $W$ and (b) on-site disorder strength $V$. The parameters of the pump loop are chosen as $\{R_\Delta,R_\delta,\delta_s\}=\{1,0.4,0.1\},\{0.7,0.7,0\}$ for (a) and (b), respectively. The real-time simulation is performed to compute the pumped charge of the lattice of $L=80$ and time period $T=400$. Other parameters are the same as those in Fig. 1 in the main text.
}
\label{figS1}
\end{figure}

The effect of on-site random disorders on topological pumps has been theoretically studied in Refs.  \cite{Qin2016,Wauters2019,Hayward2021}. Here we numerically show the robustness of topological pumps against weak and moderate quasiperiodic disorders for single pump loops. Figures \ref{figS1}(a) and \ref{figS1}(b) show numerical results of the Chern number $C$ and pumped charge $\Delta Q$ as functions of the hopping disorder strength $W$ and on-site disorder strength $V$, respectively. One can find that the quantized Chern number $C=1$ and pumped charge $\Delta Q=1$ preserve from the clean limit to finite disorder regime. For strong disorder, the Thouless pumping breaks with non-quantized pumped charge, similar as the breakdown of Thouless pumping driven by  on-site random disorders \cite{Qin2016,Wauters2019,Hayward2021}.

\begin{figure}[!h]
	\centering
	\includegraphics[width=0.46\textwidth]{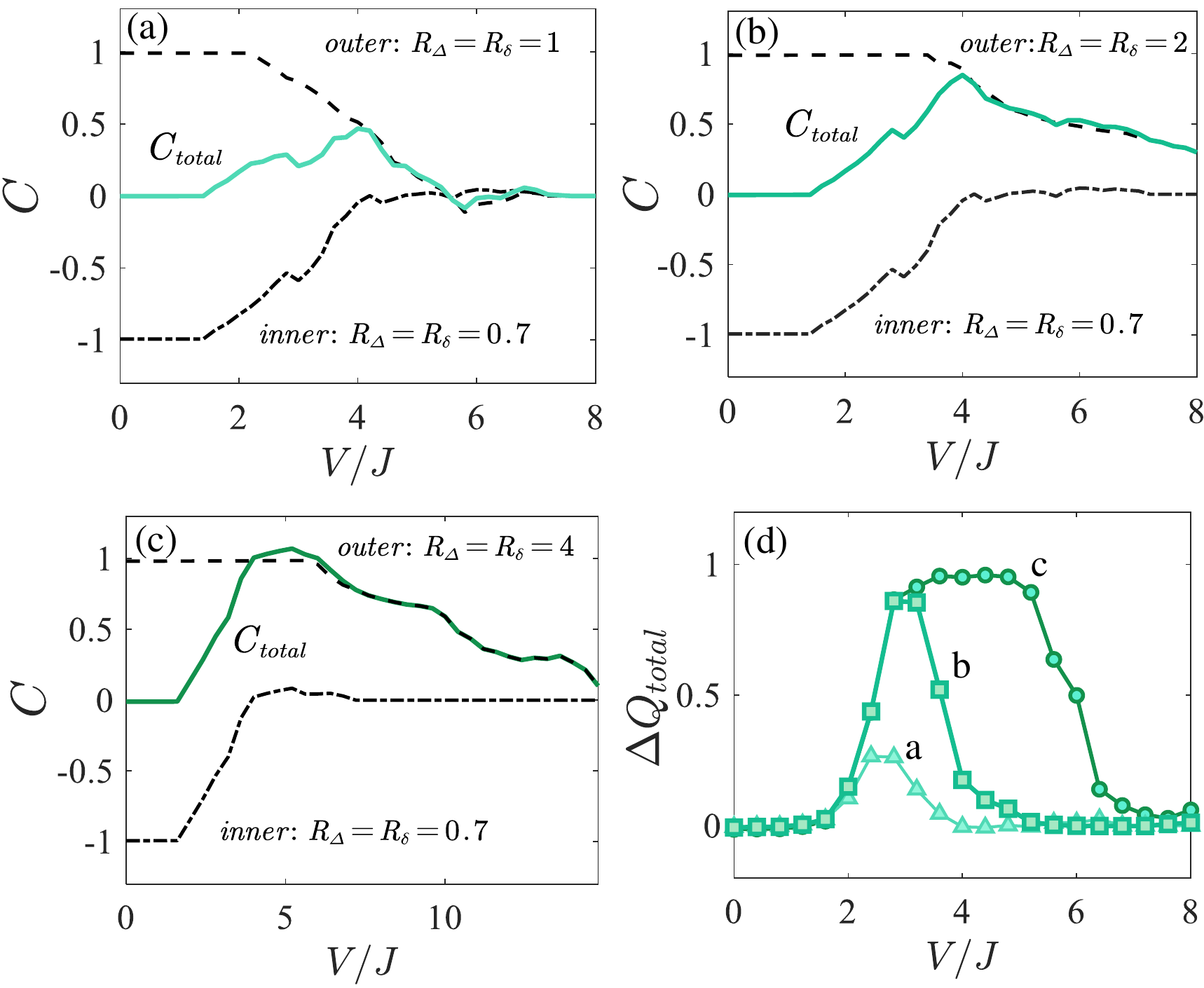}
	\caption{(Color online) $C$ as a function of the on-site disorder strength $V$ in a pump sequence connecting two loops with fixed inner loop $R_\Delta= R_\delta=0.7$ and varying outer loops (a) $R_\Delta=R_\delta=1$, (b) $R_\Delta=R_\delta=2$, and (c) $R_\Delta=R_\delta=4$. (d) Total pumped charge $\Delta Q_{total}$ as a function of $V$ for corresponding loops in (a-c). Other parameters are the same to those in Fig. 1 in the main text.
	}
	\label{figS2}
\end{figure}

\begin{figure*}[t]
\centering
\includegraphics[width=0.8\textwidth]{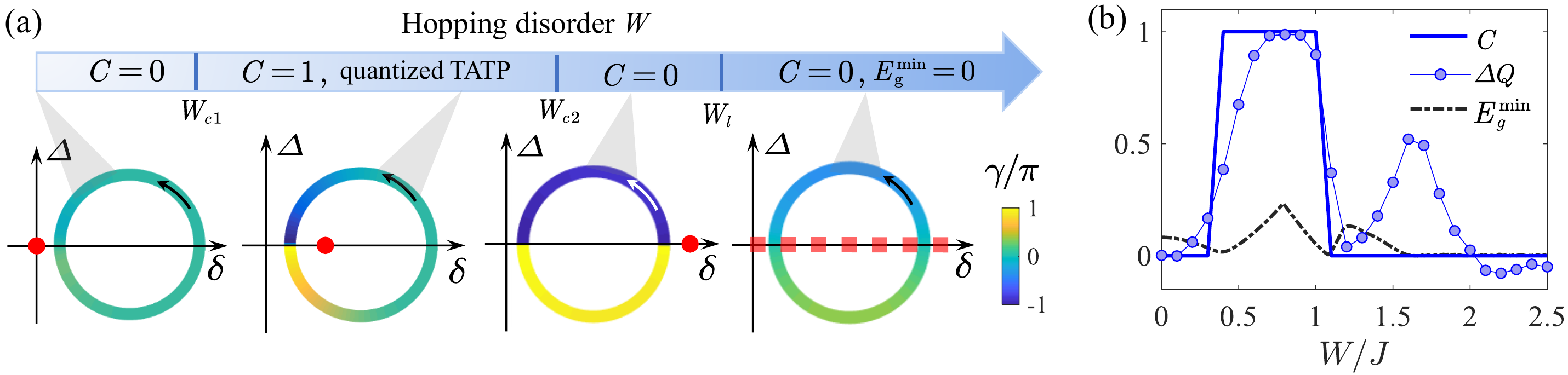}
\caption{(Color online) (a) Schematic of topological pumping induced by hopping quasiperiodic disorders. The pumping cycles, with colors and directions denoted by the Berry phase $\gamma$ and the arrows, are sketched in the $\Delta$-$\delta$ plane. The red points and dashed lines denote gapless regimes. The critical disorder strengths are $\{W_{c1},W_{c2},W_l\}\approx \{0.4,1.0,2.0\}$. Pump loops shown from left to right are $W=\{0,0.8,1.3,2.3\}$. (b) The Chern number $C$, pumped charge $\Delta Q$ and the minimum energy gap $E_g^{min}$ as a function of $W$. Parameters for computing $C$ are $L=40$ and for $\Delta Q$ are $\{L = 80, T=800\}$. Parameter of loop are chosen as $\{R_\Delta,R_\delta,\delta_s\}=\{1,0.08,0.1\}$, and other parameters are the same to those in Fig. 1 in the main text.
}
\label{figS3}
\end{figure*}

\subsection{B. Non-quantized and quantized pumping induced by on-site disorders}

Taking advantage of the robustness and breakdown of Thouless pumps, the on-site disorder-induced topological pumping can be designed by two proper connecting loops of opposite pump directions. However, this scheme requires that the two loops are apart from each other far enough, as the case shown in Fig. 1 in the main text. For inappropriately choose of inner or outer loops, the total pumped charge will become non-quantized. Here we further illustrate this point for the two-loop pump sequence. In the Fig. \ref{figS2}, we show the loop-parameter ($R_{\Delta}$-$R_{\delta}$ here) dependence of the Chern number $C$ and corresponding total pumped charge $\Delta Q_{total}$ for three different pump sequences. Here we fix the inner loop and vary the outer loop for illustration. When the outer loop is too close to the inner one, as shown in Fig. \ref{figS2}(a,b), the total Chern number $C_{total}$ takes non-integer values in the moderate disorder regime. This is due to the imperfect cancellation between opposite Chern numbers for inner and outer loops. For the large outer loop, the quantized total Chern number exhibit in the moderate disorder regime, as shown in Fig. \ref{figS2}(c) and Fig. 1 in the main text. The total pumped charges corresponding to the three pump sequences are obtained from real-time simulations and plotted in Fig. \ref{figS2}(d). The simulation results demonstrate the non-quantized and quantized charge pumping induced by the on-site disorders in (a,b) and (c), respectively.

\begin{figure}[!h]
	\centering
	\includegraphics[width=0.48\textwidth]{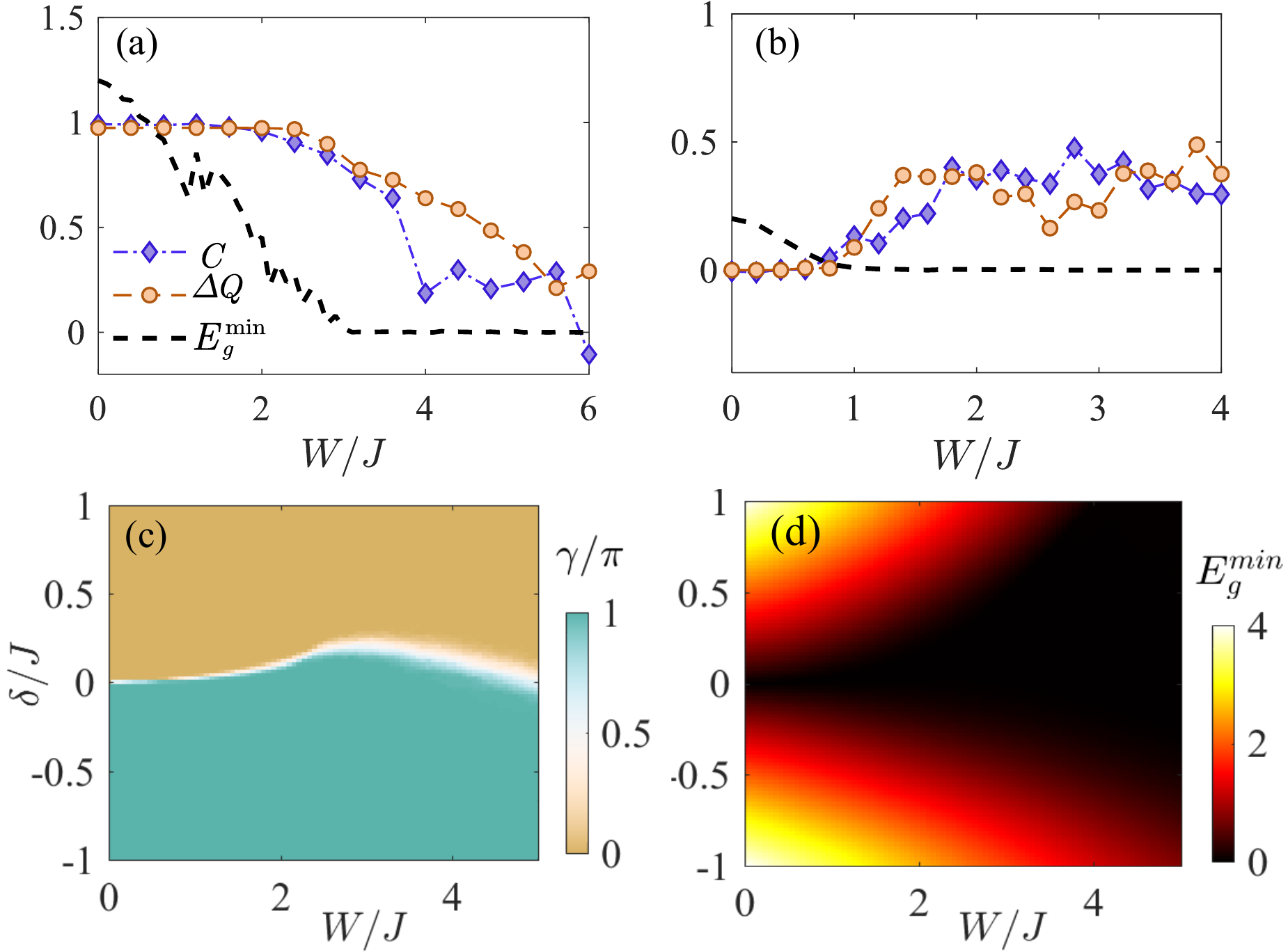}
	\caption{(Color online) $C$, $E^{min}_g$ and $\Delta Q$ as a function of the random disorder strength $W$ for the pump loop in the $\Delta$-$\delta$ plane with parameters (a) $\{R_\Delta,R_\delta,\delta_s\}=\{1,0.4,0.1\}$ and (b) $\{R_\Delta,R_\delta,\delta_s\}=\{1,0.45,0.5\}$.  All data are averaged over 20 random samples. (c) $\gamma$ and (d) $E^{min}_g$ as functions of $\delta$ and $W$ with fixed $\Delta=0$. Other parameters are $L=40$, $T=200$ and $J=1$.}
	\label{figS4}
\end{figure}

\subsection{C. Another case of breakdown of TATP}

Here we supplement another case of breakdown of TATP of specified pump loops with small $R_\delta$ and $\delta_s$. As shown in Fig. \ref{figS3}(a), by increasing the disorder strength from the clean regime, the gapless point moves along the $\delta$ axis and enters into the pump loop. This leads to the transition from the trivial pump to the TATP in the moderate disorder regime $W_{c1}<W<W_{c2}$. For the pump loop in the main text, the gapless point inside the loop becomes gapless line with $W_l<W_{c2}$ and finally the gapless line crosses the pump when $W>W_{c2}$. However, for the small loop cycle in this case, the gapless point moves out of the loop and thus breaks the TATP without closing the energy gap. Further increasing $W$, the outside gapless point becomes gapless line when $W_l>W_{c2}$ and finally crosses the trivial loop. In Fig. \ref{figS3}(b), we show the Chern number $C$, pumped charged $\Delta Q$ and minimum gap $E^{min}_g$ as a function of $W$ for such a pump. As excepted, the quantized TATP exhibits in the region $W_{c1}<W<W_{c2}$. It breaks down with gapped nature in the regime $W_{c2}<W<W_l$. In the region $W>W_l$ with $E^{min}_g=0$, non-quantized pumped charge exhibits due to the adiabaticity breakdown.

\subsection{D. Topological pumping under random disorders}

Here we show that the TATP induced by the quasiperiodic disorder for single pump loops is absent under random disorders in our model. For comparisons, we consider the same Rice-Mele Hamiltonian in Eq. (1) in the main text and replace the quasiperiodic disorder $W_j$ with the random disorder uniformly distributed in $[-W,W]$. In this random disorder case, we numerically compute the corresponding Chern number $C$, pumped charge $\Delta Q$, and minimum energy gap $E^{min}_g$ for two typical pump cycles in Figs. \ref{figS4}(a,b). For a topological pump in Fig. \ref{figS4}(a), the quantized charge pumping with $C=\Delta Q=1$ is robust against weak and moderate random disorder, which is similar to the quasiperiodic case. However, we find that the random disorder can not induce the TATP in this pump sequence. As shown in Fig. \ref{figS4}(b), the obtained $C$ and $\Delta Q$ vary from zero in the clean regime to non-quantized values when increasing the random disorder strength $W$. This is due to the fact that small or moderate random disorder [when $W>1$ in Fig. \ref{figS4}(b)] induces energy gap closing before moving the gapless point inside the pump loop.
Furthermore, we numerically check the failure of random-disorder-induced TATP for different pump loops in large parameter space. For instance, in Figs. \ref{figS4}(c,d), we show the Berry phase $\gamma$ and minimum gap $E^{min}_g$ on the $W$-$\delta$ plane with fixed $\Delta=0$. One can find that the random disorder strength $W$ can modify the critical regime ($\delta_c$ for given small $W$) between the topological and trivial phases ($\gamma=\pi,0$, respectively). However, the  regime quickly becomes line along the $\delta$ axis for given moderate $W$. Thus the TATP does not exhibit in the moderate random disorder regime.

\subsection{E. Finite-size scaling and configuration-independence of TATP}

\begin{figure}[t]
	\centering
	\includegraphics[width=0.3\textwidth]{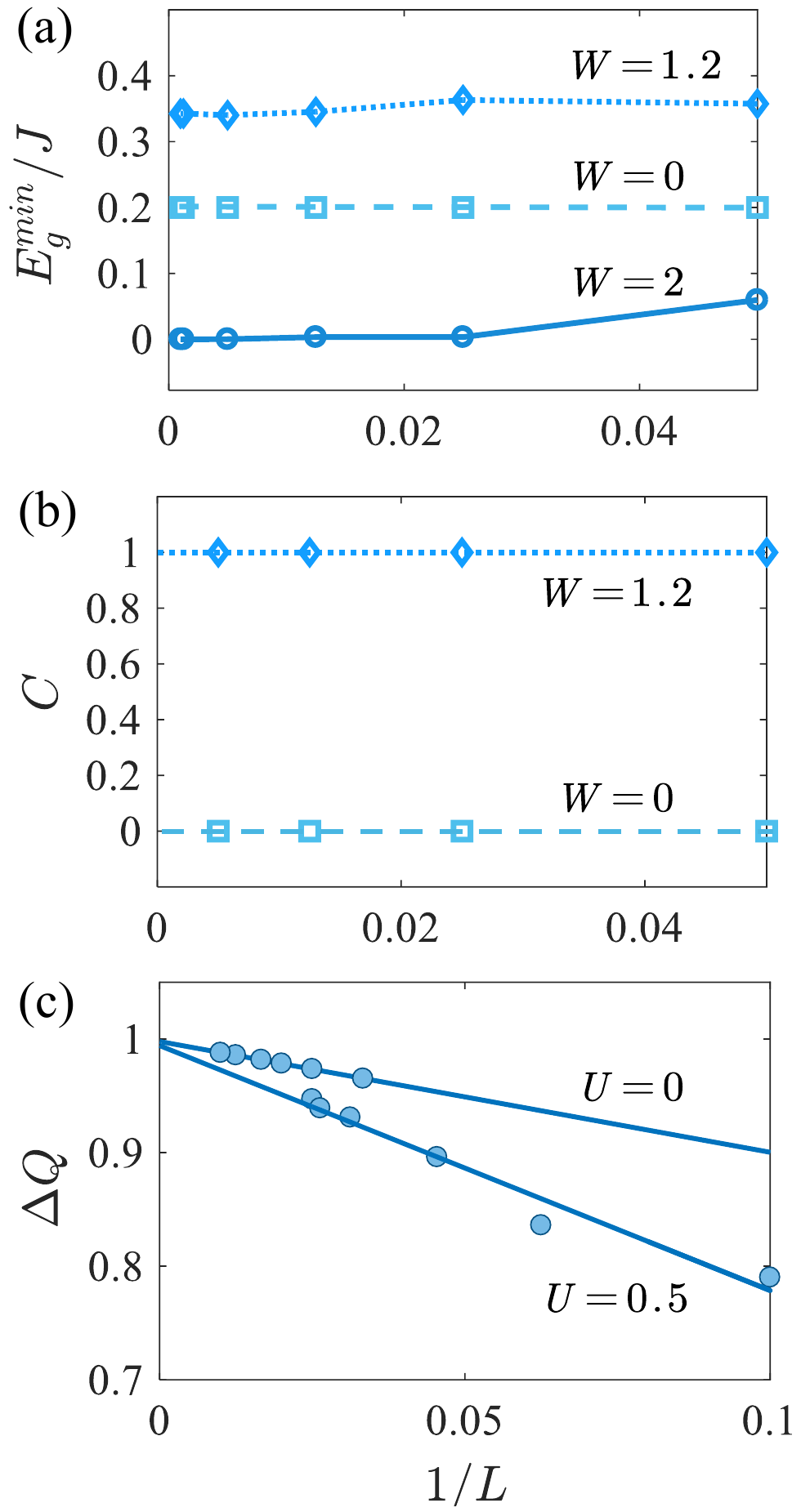}
	\caption{(Color online) Finite-size scaling of (a) minimum energy gap $E^{min}_g$, (b) Chern number $C$, and (c) pumped charge $\Delta Q$. The parameters are $\{R_\Delta, R_\delta, \delta_s \}$ = $\{1, 0.45.0.5\}$, $\beta=0$, $J=1$ in (a-c), and $U=0$ in (c). The disorder strength is $W=1.2$ in (c) and the solid lines denote the corresponding linear fittings. }
	\label{figS5}
\end{figure}

In the main text, we numerically compute the minimal energy gap $E^{min}_g$ for $L=800$, Chern number $C$ and pumped charge $\Delta Q$ for $L=40$ (or $L=80$). Here we show our numerical results of the noninteracting and interacting TATPs preserves in the large $L$ limit with negligible finite-size effects. To this end, we perform the finite-size scaling of $E^{min}_g$, $C$, and $\Delta Q$ in Figs. \ref{figS5}(a-c), respectively. Figure \ref{figS5}(a) displays the scaling of $E^{min}_g$ for two trivial pumps ($W=0$ and $W=2$) and the TATP ($W=1.2$) in Fig. 1(b) in the main text. The results show their gapped or gapless nature in the thermodynamic limit $L\rightarrow \infty$. Figure \ref{figS5}(b) shows the corresponding Chern numbers preserve for various $L$ without the finite-size effect. Moreover, we compute the scaling of $\Delta Q$ from the real-time simulations for noninteracting ($U=0,~W=1.2$) and interacting ($U=0.5,~W=1.2$) TATPs in Fig. \ref{figS5}(c). The corresponding linear fittings indicate the quantized pumped charge $\Delta Q\rightarrow1$ in the limit $L\rightarrow \infty$ in the two TATPs.

\begin{figure}[!h]
\centering
\includegraphics[width=0.48\textwidth]{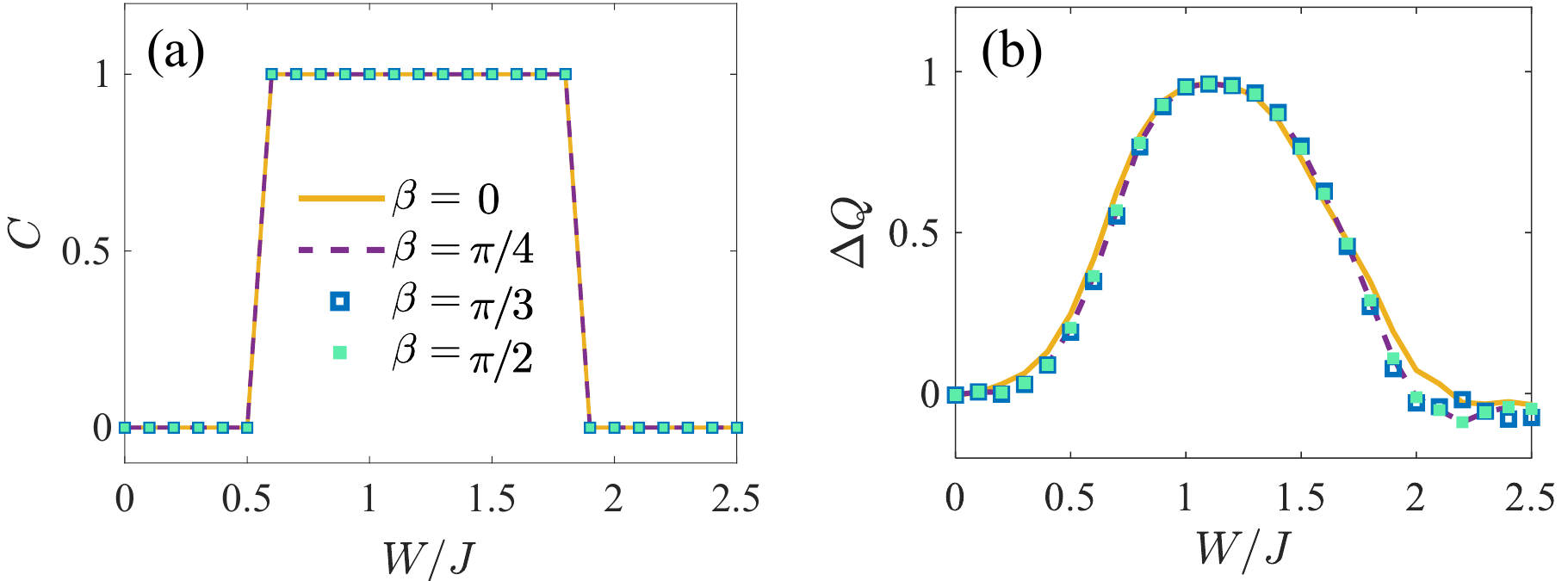}
\caption{(Color online) (a) Chern number $C$ and (b) pumped charge $\Delta Q$ as a function of hopping disorder strength $W$ for different quasiperiodic lattice configurations with $\beta=0,\pi/4,\pi/3,\pi/2$. Other parameters are $\{R_\Delta, R_\delta, \delta_s \}$ = $\{1, 0.45.0.5\}$, $L=40$, $T=200$, $U=0$, and $J=1$.}
\label{figS6}
\end{figure}

In the main text, we focus on the quasiperiodic lattice configuration with $\beta=0$. In Fig. \ref{figS6}, we show that our main results of $C$ and $\Delta Q$ for exhibiting the TATP are independent of the configurations. This can be clearly seen from Fig. \ref{figS6} that the results of $C$ and $\Delta Q$ as a function of $W$ are almost $\beta$-independent. Thus, one can use a quasiperiodic lattice configuration to realize and detect the TATP in experiments.

\subsection{F. Localization properties of adiabatic eigenstates}

\begin{figure*}[t]
\centering
\includegraphics[width=0.75\textwidth]{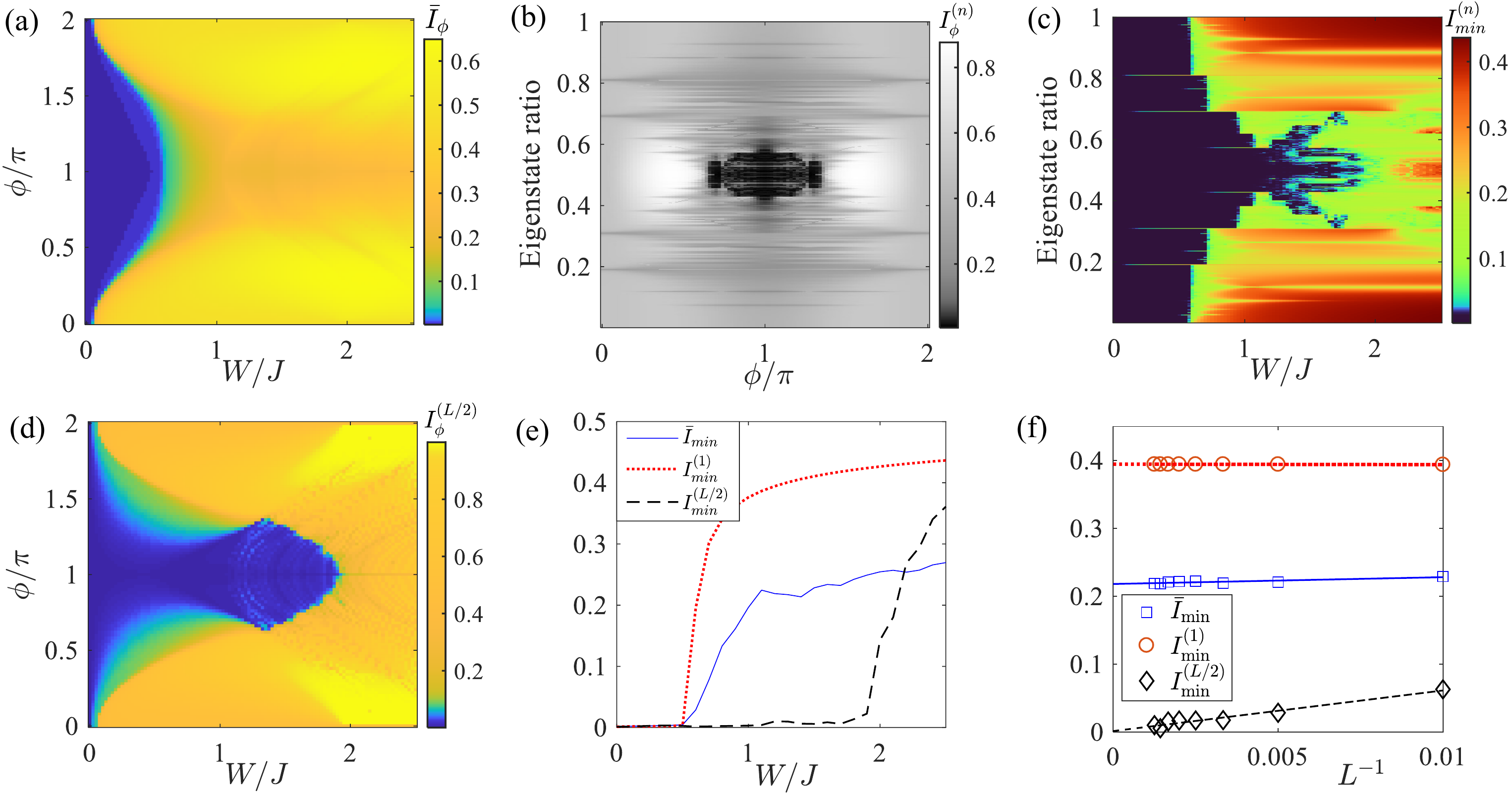}
\caption{(Color online) (a) The averaged IPR $\bar{I}_\phi$ as functions of quasiperiodic disorder strength $W$ and the pump parameter $\phi$. (b) The IPR $I^{(n)}_\phi$ as functions of $\phi$ and the eigenstate ratio (i.e., $n/L$). (c) The minimum IPR $I^{(n)}_{\min}$ as functions of $W$ and the eigenstate ratio. (d) $I^{(L/2)}_\phi$ as functions of $W$ and $\phi$. (e) $\bar{I}_{\min}$ (blue solid line), $I^{(1)}_{\min}$ (red dotted line) and $I^{(L/2)}_{\min}$ (black dashed line) as a function of $W$. (f) Finite-size scaling of $\bar{I}_{\min}$, $I^{(1)}_{\min}$ and $I^{(L/2)}_{\min}$ for $W=1.2$. Other parameters are $\{R_\Delta, R_\delta, \delta_s \}$ = $\{1, 0.45.0.5\}$, $L=800$, $U=0$, and $J=1$.}
\label{figS7}
\end{figure*}

The localization effect on topological pumps from the on-site random disorders has been studied in Refs. \cite{Wauters2019,Hayward2021}. In this part, we numerically study the localization properties of instantaneous eigenstates during the adiabatic pump sequence under the PBC. For $n$-th normalized eigenstate $|\psi^{(n)}_\phi\rangle$ for the adiabatic pump parameter $\phi$ (and disorder strength $W$), the inverse participation ratio (IPR) is defined as \cite{YXLiu2020,CMDai2018,RoyShilpi2021,YTHsu2018,SLXu2019,XPLi2016}
\begin{equation}
	I^{(n)}_\phi=\sum_{j=1}^{L}\left|\langle j|\psi^{(n)}_\phi\rangle\right|^{4}.
\end{equation}
For the large lattice size $L$, one has $I^{(n)}_\phi\sim \mathcal{O}(1)$ for a localized ($n$-th) eigenstate, while $I^{(n)} _\phi \sim {L^{-1}}$ for an extended eigenstate. The overall localization property of all the eigenstates can be characterized by the averaged IPR $\bar{I}_\phi=\frac{1}{L}\sum_{n=1}^L I^{(n)}_\phi$. To reveal the eigenstate localization during the pump cycle, we can use the minimum IPR of the $n$-th eigenstate $I^{(n)}_{\min}= \min_{\phi} \{I^{(n)} _\phi \}$ and the minimum averaged IPR $\bar{I}_{\min}=\min_{\phi}\{ \bar{I}_\phi \}$, with $\phi$ varying from $0$ to $2\pi$.

We first plot the averaged IPR $\bar{I}_\phi$ as functions of the quasiperiodic disorder strength $W$ and the pumping parameter $\phi$ in Fig. \ref{figS7}(a). The numerical result for the lattice system of size $L=800$ shows that the instantaneous eigenstates adiabatic during the whole pumping progress are fully extended only for very small $W\lesssim0.1$. When $W\gtrsim0.5$, most (or all) of the eigenstates will become localized. For instance, in Fig. \ref{figS7}(b), we show the IPR of all eigenstates $I^{(n)}_\phi$ as a function of $\phi$ for the TATP with $W=1.2$. One can see clearly that most of eigenstate are localized during the whole TATP cycle, and only some eigenstates in the spectrum center (i.e., $n/L\sim0.5$) preserve extended when $0.6\pi\lesssim\phi\lesssim1.4\pi$. Figure \ref{figS7}(c) shows the minimum IPR $I^{(n)}_{\min}$ of all eigenstates during the pump cycle as a function of $W$. One can find the two middle eigenstates (i.e. the $n=L/2,L/2+1$) in the adiabatic energy spectrum are the last ones to become localized with increasing $W$, which occurs when $W\gtrsim1.9$. Moreover, we show $I^{(L/2)}_{\min}$ of the $L/2$-th eigenstate as functions of $W$ and $\phi$ in Fig. \ref{figS7}(d). The result indicates that all eigenstates during the entire pump cycle will become localized when $W\gtrsim1.9$ in the trivial pump regime. To show the above results more clearly, one also plot $\bar{I}_{\min}$, $I^{(1)}_{\min}$ and $I^{(L/2)}_{\min}$ as a function of $W$ in Fig. \ref{figS7}(e). Finally, we note that these numerical results for the lattice of size $L=800$ preserve in the large $L$ limit, with the corresponding finite-size scaling for $W=1.2$ shown in Fig. \ref{figS7}(f).

\subsection{G. Numerical methods of DMRG and TDVP}

The density matrix renormalization group (DMRG) and time-dependent variational principle (TDVP) methods are very useful for numerical calculations of static and dynamical properties of 1D strongly correlated systems \cite{White1992,Schollwoeck2011}. Here we provide some technical details of the DMRG and TDVP methods, and benchmark the numerical results for our model. Given a general 1D quantum state $\ket{\Psi}$ of the many-body Hamiltonian $\hat{H}$, one can obtain its matrix product state (MPS) representation as \cite{White1992,Schollwoeck2011}
\begin{equation}
	\begin{aligned}|\Psi\rangle=& \sum_{j_{1}, \ldots, j_{L}} \sum_{a_{0}, a_{1}, \ldots, a_{L}} \operatorname{Tr}\left(M_{a_{0}, a_{1}}^{j_{1}} M_{a_{1}, a_{2}}^{j_{2}} \cdots\right.\\ &\left.\times M_{a_{L-2}, a_{L-1}}^{j_{L-1}} M_{a_{L-1}, a_{L}}^{j_{L}}\right)\left|j_{1}, j_{2}, \ldots, j_{L-1}, j_{L}\right\rangle. \end{aligned}
\end{equation}
Here $M_{a_{n-1},a_n}^{j_n}$ are the rank-three tensors with $a_n$ denoting a virtual index which links adjacent sites, and $j_n$ is the index of a local state. The indices of $a_0,a_L$ are dummy under the OBCs and the operation of trace is needless. Using the gauge freedom, one can obtain the canonical form of MPS:
\begin{equation}
	|\Psi\rangle=\sum_{\alpha=1}^{\chi_{n}} \Lambda_{\alpha \alpha}^{[n]}\left|l_{\alpha}^{[n]}\right\rangle\left|r_{\alpha}^{[n]}\right\rangle,
\end{equation}
where $\Gamma^{[n]}$ stands a diagonal matrix including the Schmidt values of the bi-partition $\{l,r\}$, $n$ is the canonical center, $\chi$ is the total amount of the Schmidt values, the orthogonal Schmidt states $\left|l_{\alpha}^{[n]}\right\rangle=\sum_{j_{1}, \ldots, j_{n}}\left(A^{j_{1}} \cdots A^{j_{n}}\right)_{\alpha}\left|j_{1}, \ldots, j_{n}\right\rangle$ and $\left|r_{\alpha}^{[n]}\right\rangle=\sum_{j_{n+1}, \ldots, j_{L}}\left(B^{j_{n+1}} \cdots B^{j_{L}}\right)_{\beta}\left|j_{n+1}, \ldots, j_{L}\right\rangle$ corresponds to the left and right parts, respectively, and $A^{j_i}$ and $B^{j_i}$ are the left and right canonical form of $M^{j_i}$ at site $i$. By optimizing two linked MPS tensors $M^{j_{n}, j_{n+1}}=A^{j_{n}} \Lambda^{[n]} B^{j_{n+1}}$ and minimizing the ground state energy $\bra{\Psi_0}\hat{H}\ket{\Psi_0}$, the two-site DMRG methods will converge to the ground state $\ket{\Psi_0}$. The total algorithm can be abstracted as follows \cite{Schollwoeck2011}:

(1) Prepare the MPS wave function $\Psi_0$ whose canonical center is site $n$:
\begin{equation}
	\left|\Psi_{0}\right\rangle=\sum_{\alpha, j_{n}, j_{n+1}, \beta} M^{j_{n}, j_{n+1}}\left|l_{\alpha}^{[n-1]}\right\rangle\left|j_{n}\right\rangle\left|j_{n+1}\right\rangle\left|r_{\beta}^{[n+1]}\right\rangle.
\end{equation}
In each variational update, the tensors belonging to $\ket{l^{[n-1]}}$ and $\ket{r^{[n+1]}}$ will be fixed and $M^{j_{n}, j_{n+1}}=A^{j_{n}} \Lambda^{[n]} B^{j_{n+1}}$ should be improved.

(2) Obtain the effective Hamiltonian $\hat{H}_{eff}$ in the projected basis $\ket{l_\alpha^{[n-1]} j_n j_{n+1} r_\beta^{[n+1]}}$.

(3) Numerically obtain the lowest-lying eigenvector $\tilde{M}^{j_n,j_{n+1}}$ of the effective Hamiltonian $\hat{H}_{eff}$.

(4) Sweep from left: update tensor $A^{j_n}=\tilde{A}^{j_n}$ on sites $n$ by applying singular value decomposition (SVD) to $\tilde{M}^{j_n,j_{n+1}}=\tilde{A}^{j_n} \tilde{\Lambda}^{[n]} \tilde{B}^{j_{n+1}}$; prepare tensor $M^{j_{n+1},j_{n+2}}=\tilde{\Lambda^{[n]}} \tilde{B}^{j_{n+1}} B^{j_{n+2}}$ for the next sites. Sweep from right: update tensor $B^{j_{n+1}}=\tilde{B}^{j_{n+1}}$ on sites $n+1$ by applying singular value decomposition (SVD) to $\tilde{M}^{j_n,j_{n+1}}=\tilde{A}^{j_n} \tilde{\Lambda}^{[n]} \tilde{B}^{j_{n+1}}$; prepare tensor $M^{j_{n-1},j_{n}}=  A^{j_{n-1}} \tilde{A}^{j_{n}} \tilde{\Lambda^{[n]}}$ for the next sites.

(5) Return to the step 2 and and sweep repeatedly till convergence.

\begin{figure}[t]
\centering
\includegraphics[width=0.48\textwidth]{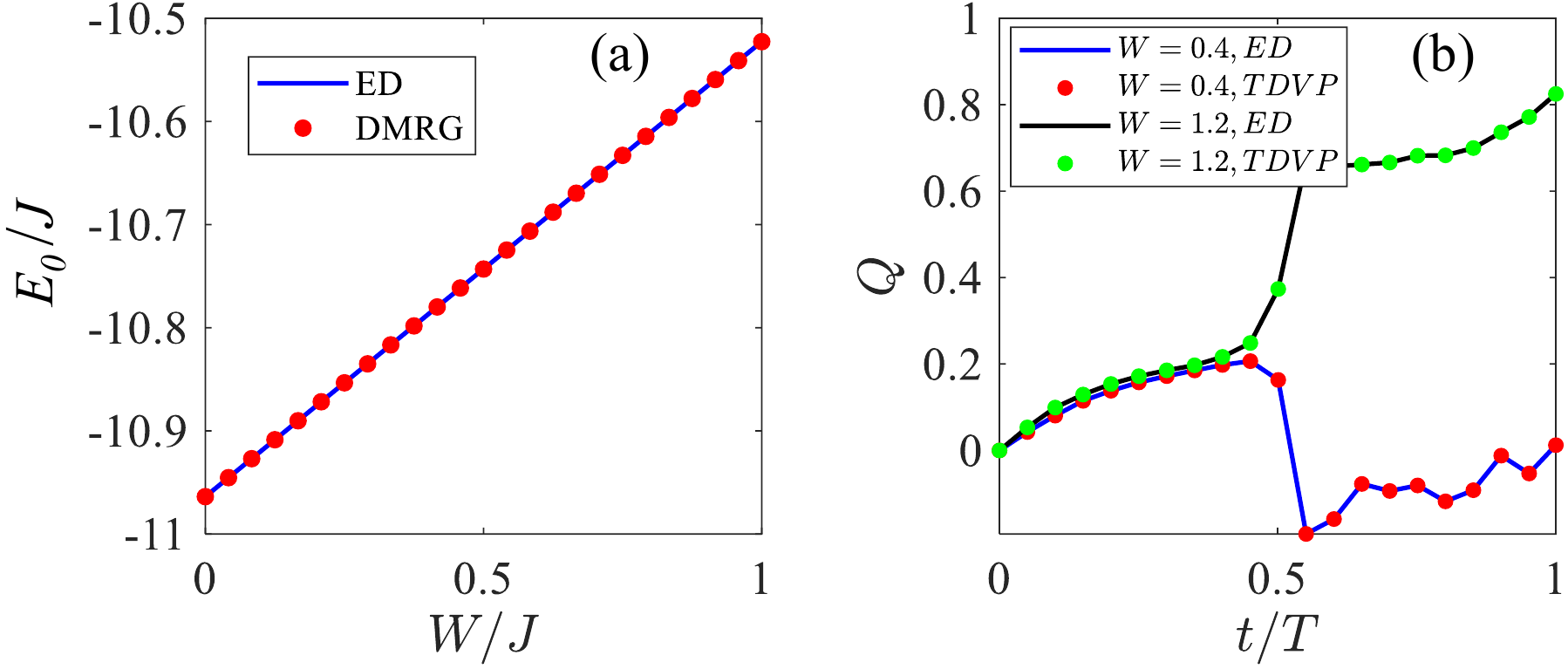}
\caption{(Color online) (a) Ground state energy $E_0$ as a function of the hopping disorder strength $W$ calculated by ED (blue solid line) and DMRG (red point). (b) Time evolution of the pumped charge $Q(t)$ calculated by ED (blue solid line for $W=0.4$ and black solid line for $W=1.2$) and TDVP (red point for $W=0.4$ and green point for $W=1.2$). Other parameters are $\{R_\Delta, R_\delta, \delta_s \}$ = $\{1, 0.45.0.5\}$, $L=12$, $T=200$, $U=0.5$, and $J=1$.}
\label{figS8}
\end{figure}

In Fig. \ref{figS8}(a), we benchmark the numerical result of the ground state energy $E_0$ as a function of the hopping disorder strength $W$ with fixed $U=0.5$, by using methods of the exact diagonalization (ED) and the DMRG. To obtain the time evolution of the many-body wave function $\ket{\Psi(t)}$ in Eq. (2) in the main text (when $U\neq0$), one can numerically compute $\ket{\Psi(t)}=\hat{\mathcal{T}}\exp(-i\int_0^t\hat{H}(t') dt') \ket{\Psi(0)}$. This can be achieved by the ED when the system size $L\leq16$ in our numerical calculations at half filling. However, the ED fails when $L>16$ due to the large calculation amount. In this case, we use the TDVP to compute the pumping dynamics in the presence of interactions and disorders.

The TDVP method is the same as the DMRG method except the third implementation step, where we should replace the eigen-solver by applying the matrix exponential on the local wave function: $\tilde{M}^{j_{n}, j_{n+1}}=\exp \left(-i \hat{H}_{\mathrm{eff}} \tau\right) M^{j_{n}, j_{n+1}}$ with the time step $\tau$ \cite{Haegeman2016}. There is an additional operation after the SVD in step (4), backward evolution of $\tilde{A}^{j_{n}} \tilde{\Lambda}[n]$ (while sweeping from right) or $\tilde{\Lambda}^{[n]} \tilde{{B}}^{j_{n+1}}$ (while sweeping from left) before preparation of the tensor $M^{j_{n+1}, j_{n+2}} \text { or } M^{j_{n-1}, j_{n}}$. This operation is actually the same as the third step with a small time step $-\tau$. The targeted wave function $\ket{\Psi(\tau)}$ can be obtained by sweeping from $n=1$ to $n=L-1$. We can repeat the sweep to obtain long-time evolved wave function.

In Fig. \ref{figS8}(b), we benchmark the numerical result of the time evolution of the pumped charge $Q(t)$ for different disorder strength $W$ with fixed $U=0.5$, by using methods of ED and TDVP, respectively. Noted that the obtained pumped charge $\Delta Q=Q(T)-Q(0)$ is not quantized for $W=1.2$ due to small system size. By using the TDVP, we compute $\Delta Q$ up to $L=40$ at half filling as shown in Fig. 2(d) in the main text and Fig. \ref{figS5}(c) (with the finite-size scaling).

\subsection{H. Higher-order topological pump in 2D systems}
Here we provide more details on the higher-order topological pumping in 2D systems \cite{Wienand2022,Benalcazar2022b,BXie2021}. As described in the main text, we focus on a 2D super-lattice (of size $L\times L$) described by the following Hamiltonian
\begin{equation}
	\begin{aligned}
		\hat{H}_{2 D}=&-\sum_{x, y=1}^{L}\left[J_{x} \hat{c}_{x, y}^{\dagger} \hat{c}_{x+1, y}+(-1)^{x} J_{y} \hat{c}_{x, y}^{\dagger} \hat{c}_{x, y+1}+\text {H.c.}\right] \\
		&+\Delta \sum_{x, y=1}^{L}(-1)^{x+y} \hat{n}_{x, y}.
	\end{aligned}
\end{equation}
When the staggered potential $\Delta=0$ and the hopping strengths $J_{x,y}$ are staggered along the two dimensions, the Hamiltonian describes the so-called higher-order topological insulators (HOTIs), as a 2D extension of the SSH model \cite{Benalcazar2017a,Benalcazar2017b,FLiu2017}. The 2D HOTIs are characterized by a quantized electric quadrupole moment in the bulk and zero-energy modes with fractional charges at the corners. In the presence of (random or quasiperiodic) hopping disorders, all the crystalline symmetries are broken, but the (sublattice) chiral symmetry preserves when $\Delta=0$. It has been revealed that, as long as the chiral symmetry is preserved, the HOTI phase can exhibit and even induced by the disorder \cite{CLi2020,YBYang2021,WZhang2021,Coutant2020,Benalcazar2022}. In the main text, we consider staggered hopping amplitudes in two spatial dimensions $\eta=x,y$ $J_\eta=J\equiv1$ and $J_\eta=m_\eta=J+\delta+W\cos(2\pi\alpha \eta)$ with quasiperiodic disorder $W$. In the clean HOTI phase ($W=\Delta=0$) with $\delta<0$, the ground state at half-filling under the OBC takes quantized fractional corner charge $q_c=\pm {1}/{2}$, while $q_c=0$ in the trivial phase with $\delta>0$. The topological phase transition point will be driven from $\delta=0$ in the clean regime to other values of $\delta$ by moderate disorder strengths, which leads to the higher-order topological Anderson insulators \cite{CLi2020,YBYang2021,WZhang2021}.

To realize the higher-order Thouless pumping in this 2D system \cite{Wienand2022,Benalcazar2022b,BXie2021}, we have to add the staggered potential $\pm\Delta$ to break the chiral symmetry and connect the (clean or disordered) HOTI and trivial phases. Here we consider the staggered potential in a cross-diagonal arrangement. A pump cycle in this 2D system can be achieved via the adiabatic parameter $\phi$ from $\delta=R_{\delta}\cos\phi+\delta_s$ and $\Delta=R_{\Delta}\sin\phi$, which form a closed loop on the $\delta$-$\Delta$ parameter plane when $\phi$ varies from $0$ to $2\pi$. To characterize the topology of the system under disorders and adiabatic pumping, following Ref. \cite{Wienand2022}, we use the higher-order Berry phase by introducing the corner PBC (see Fig. 3(a) in the main text). To apply the corner PBC \cite{Wienand2022}, we attach the corner-connecting link described by the following Hamiltonian
\begin{equation}
	\hat{H}_c=-J\left(
	 e^{i\theta} \hat{c}_{c_{1}}^{\dagger} \hat{c}_{c_{2}}
	+e^{-i\theta} \hat{c}_{c_{2}}^{\dagger} \hat{c}_{c_{3}}
	+\hat{c}_{c_{3}}^{\dagger} \hat{c}_{c_{4}}
	+\hat{c}_{c_{4}}^{\dagger} \hat{c}_{c_{1}}
	+\text {H.c.}\right)
\end{equation}
where $c_i$ stands the coordinates of the $i$-th corner ($i=1,2,3,4$), i.e., $c_1=(1,1)$, $c_2=(1,L)$, $c_3=(L,L)$ and $c_4=(L,1)$. The total Hamiltonian can thus be obtained as $\hat{H}_{\mathrm{tot}}(\theta)=\hat{H}_{2 D}^{\mathrm{OBC}}+\hat{H}_{c}(\theta)$. Under the corner PBC, we can define the Berry phase with the ground state wave function $\ket{\psi(\theta,\phi)}$ at half filling even in the presence of disorders:
\begin{equation}
	\gamma(\phi)=\oint_{0}^{2 \pi} d {\theta}\left\langle \psi(\theta,\phi)\left|i \partial_{\theta}\right| \psi(\theta,\phi)\right\rangle.
\end{equation}
Due to the $\mathbb{Z}_{2}$ symmetry ($\hat{c}^\dagger_{x,y}\leftrightarrow\hat{c}_{x,y}$) when $\phi=\{0,\pi\}$, the Berry phase $\gamma$ is $\mathbb{Z}_{2}$-quantized, $\gamma \in \pi \mathbb{Z}$. By applying Resta's argument to higher-order systems and introducing position operator in the bulk \cite{Wienand2022}, $\gamma$ can be related to the charge transport during the pumping. A key step in the process is that the corner link flux $\theta$ can be related to current $\hat{\mathcal{J}}=\partial_{\theta} \hat{H}_{\text{tot}}(\theta)|_{\theta=0}$ pumping through a corner diagonally. Integrating these current in a Thouless pump loop yields a total change of the corner charge $\delta q_c$ related to the Berry phase \cite{Wienand2022}: $|\delta q_c| =\Delta \gamma/2\pi=1,0$ for topological and trivial pumps (see Fig. 3(b) in the main text for instances), respectively. Here $\Delta \gamma$ denotes gauge-invariant difference of the Berry phases in a pump cycle. The total amount of the charge pumping to the corner (as the corner state) under the OBC can be characterized by the Chern number $C$, which is the winding number of the Berry phase \cite{Wienand2022}:
\begin{equation}
	C=\int_{0}^{2 \pi} \frac{d \phi}{2 \pi} \partial_{\phi} \gamma=|\delta q_c|.
\end{equation}
In the thermodynamic limit, the corner-to-corner transport associated with the higher-order adiabatic topological pumping is quantized. In our numerical simulations for the finite lattices under the OBC, we count the corner charge $q_c$ by integrating up the particle density near the corner $c_1$ (or one of the other corners with the same result of $|q_c|$) in the regime of $L/3 \times L/3$ lattices, i.e., $q_c=\sum_{x,y=1}^{L/3} (\braket{\hat{n}_{x,y}}-1/2)$ with $1/2$ as the background density at half filling. Note that $q_c$ takes the fractional values $\pm 1/2$ with high accuracy if the system is in the HOTI phase with sufficiently large lattice size, such as $L=60$ in the main text.

\subsection{I. The TATP in optical systems}

In the main text, we propose the disorder-induced TATP with fermionic atoms in optical lattices. Here we show that the non-interacting TATP can also be observed in optical systems, similar to the topological Thouless pumping of light in waveguide arrays \cite{Kraus2012a,Cerjan2020,QCheng2022,WLiu2022,YKe2016}. The required Rice-Mele Hamiltonian for the light tightly confined to the coupled single-mode waveguides can be realized by modulating their refraction index and relative distances \cite{Cerjan2020,QCheng2022}. Moreover, tunable quasiperiodic disorders have been experimentally fabricated into the arrays by varying the spacing between neighboring waveguides \cite{Kraus2012a,Cerjan2020,QCheng2022,WLiu2022}. The adiabatic pump cycle can be achieved and tuned by periodically modulating both the refraction index and the waveguide spacing as a function of the propagation distance, say $z$. Here $z$ represents the evolution time of the pump cycle. For consistency, we still use the time $t$ to denote the evolution of the pump cycle in optical systems.

\begin{figure}[!h]
\centering
\includegraphics[width=0.47\textwidth]{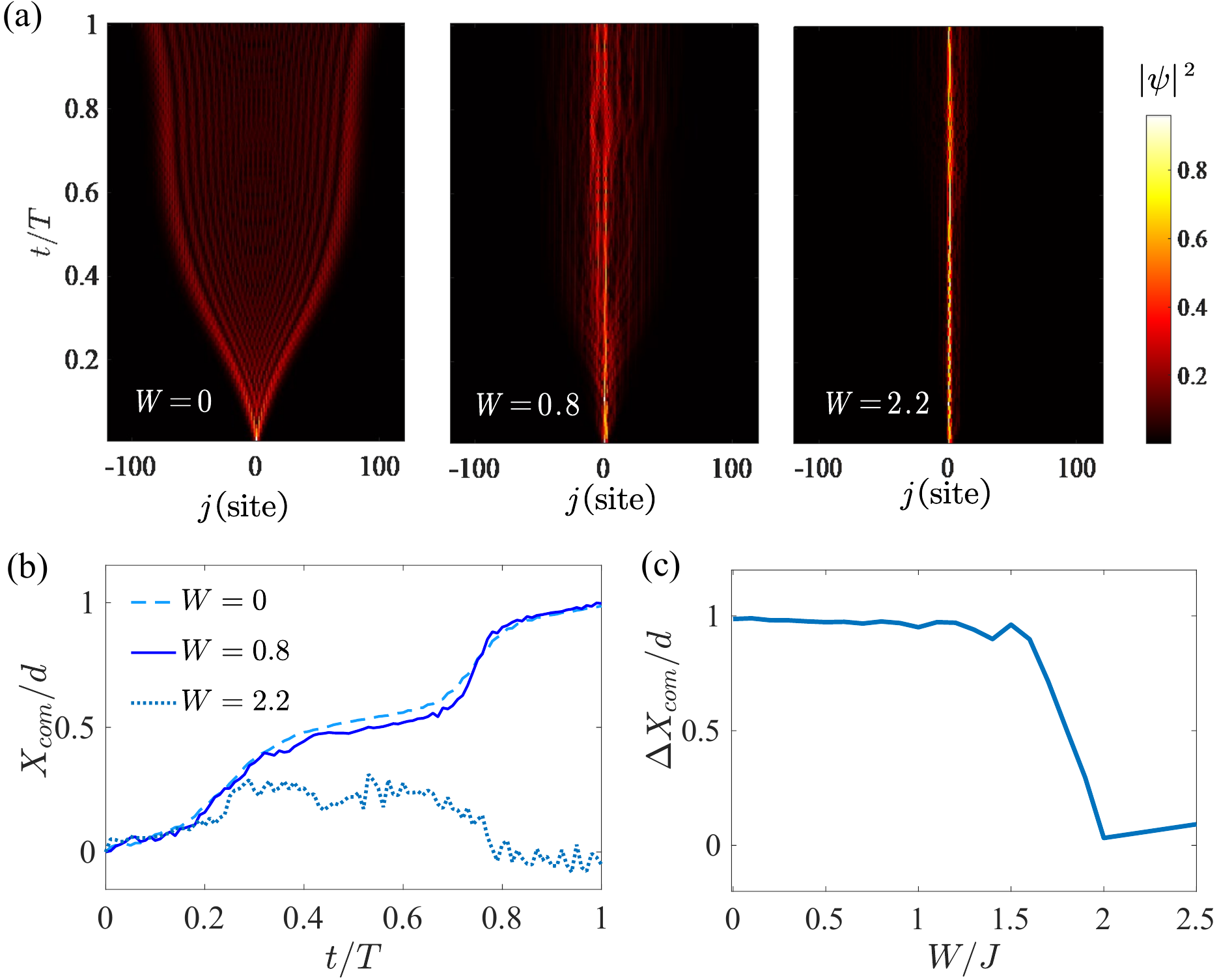}
\caption{(Color online) (a) Time evolution of the density distribution over a pump cycle for the quasiperiodic disorder strengths $W=0,0.8,2.2$, respectively. (b) Corresponding time evolution of the COM $X_{com}(t)$. (c) The COM shift $\Delta X_{com}$ as a function of $W$. The data in (b,c) are averaged over 50 (100) samples for $0<W<1.5$ ($W>1.5$). Other parameters are $\{R_{\Delta},R_{\delta},\delta_s\}=\{1,0.4,0.1\}$, $L=240$, $T=250$, and $J=1$.}
\label{figS9}
\end{figure}

\begin{figure}[!h]
\centering
\includegraphics[width=0.47\textwidth]{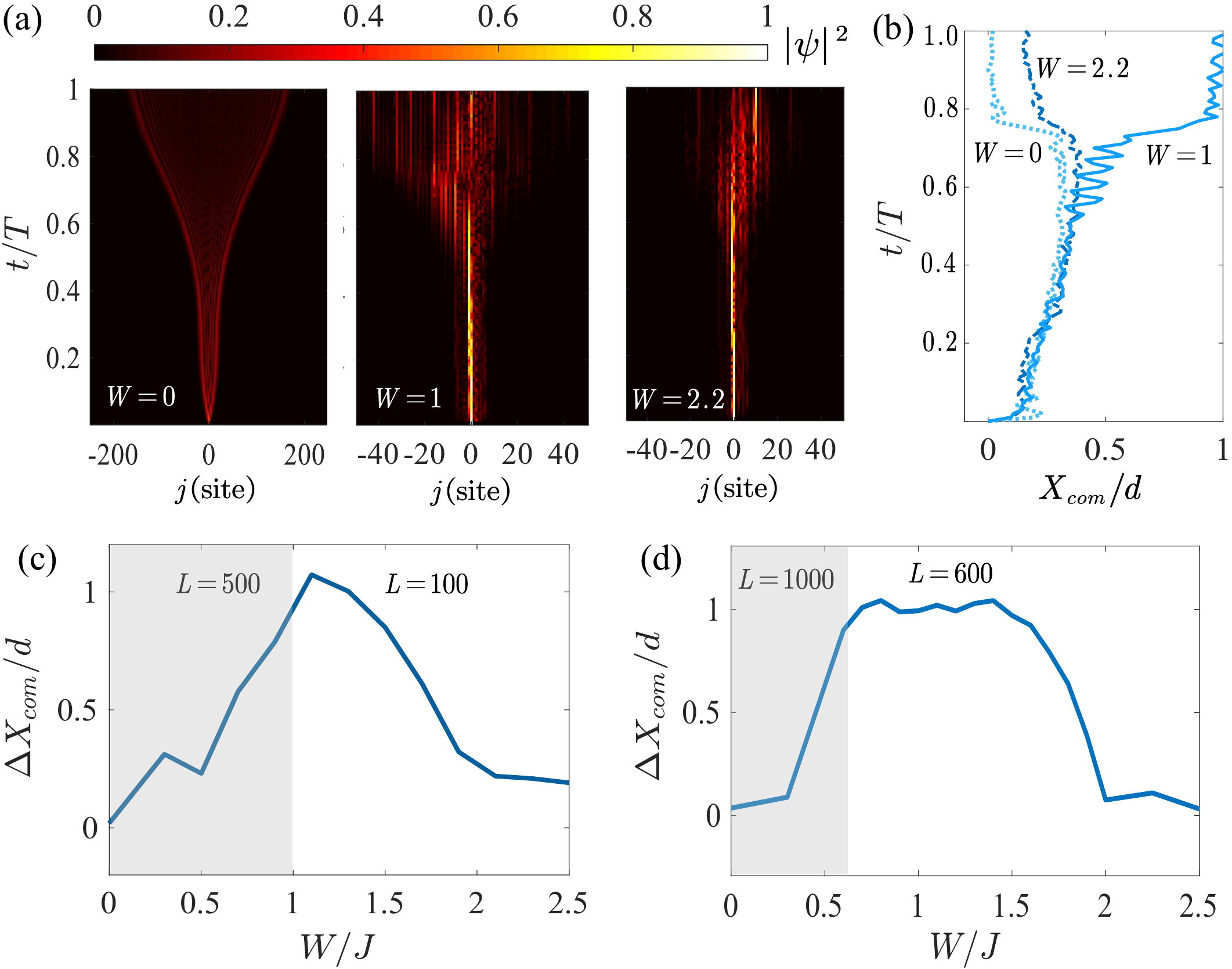}
\caption{(Color online) (a) Time evolution of the density distribution over a pump cycle for quasiperiodic disorder strengths $W=0$ ($L=500$), $W=1$ ($L=100$), and $W=2.2$ ($L=100$), respectively. (b) Corresponding time evolution of the COM $X_{com}(t)$. (c,d) The COM shift $\Delta X_{com}$ as a function of the disorder strength $W$ for different lattice sizes labeled for the corresponding disorder regimes in the figures. The data in (b-d) are averaged over 50 (100) samples for $0<W<1$ ($W>1$). Other parameters are $\{R_{\Delta},R_{\delta},\delta_s\}=\{1,0.45,0.5\}$ (the same as those in Fig. 1 in the main text), $T=250$, and $J=1$.}
\label{figS10}
\end{figure}

The pump dynamics is described by the time evolution of the wave function $|\psi(t)\rangle=\hat{\mathcal{T}}\exp(i\int_0^t\hat{H}_0(t')dt')|\psi(0)\rangle$, where $\hat{\mathcal{T}}$ is the time-ordering operator, $\hat{H}_0$ is the noninteracting Rice-Mele Hamiltonian with quasiperiodic hopping disorders in Eq. (1) in the main text. In experiments \cite{Kraus2012a,Cerjan2020,QCheng2022,WLiu2022}, the initial state of the injected light $|\psi(0)\rangle$ is prepared at one of the edges or a single site in the bulk for the pumping. Here we consider the initial state $|\psi(0)\rangle$ as the single excitation at the center of the waveguide array \cite{Cerjan2020}. Note that such an initial state is close to the (maximally) localized Wannier state and easily to be prepared in realistic optical systems \cite{Kraus2012a,Cerjan2020,QCheng2022,WLiu2022}. After injecting such a localized state into the array, one allows it to propagate to the final state $|\psi(T)\rangle$ over a full pump (modulation) cycle with the trajectory on the $\Delta$-$\delta$ plane. At the end of the cycle, one can measure the bulk transport, i.e., the center-of-mass (COM) shift $\Delta X_{com}=X_{com}(T)-X_{com}(0)$. Here the COM at time $t$ is given by $X_{com}(t)=\sum_jj\langle j|\psi(t)\rangle=\sum_jj|\psi_j(t)|^2$ with $|\psi_j|^2$ as the density at site $j$. We set the site index $j=\{-\frac{L}{2},-\frac{L}{2}+1,...,0,...,\frac{L}{2}-1,\frac{L}{2}\}$ and the initial state localized at the center site $j=0$, with $L$ large enough for the propagation over a pump cycle.

We first consider the topological pumping under disorders. In Fig. \ref{figS9}(a), we plot the time evolution of the density distribution during a pump cycle for disorder strengths $W=0,0.8,2.2$. The corresponding COM $X_{com}(t)$ in units of one unit cell of the Rice-Mele lattice $d=2a$ (with the lattice spacing $a\equiv1$) is shown in Fig. \ref{figS9}(b). One can find that the COM shift after the pump cycle $\Delta X_{com}/d\approx1$ for $W=0$ and $0.8$, while $\Delta X_{com}/d\approx0$ for $W=2.2$. Figure \ref{figS9}(c) shows $\Delta X_{com}$ as a function of $W$ obtained from real-time simulations. The result demonstrates the robustness of the topological pump with nearly quantized COM shift against disorders ($W\lesssim1.6$), which corresponds to the topological bulk transport of the single-site injected light.

The disorder-induced TATP can be observed from the COM shift of the injected light in Fig. \ref{figS10}. Figures \ref{figS10}(a) and (b) show the time evolution of the density distribution for $W=0,1,2.2$ and the corresponding COM evolution $X_{com}(t)$, respectively. The COM shift after the trivial pump cycle $\Delta X_{com}/d\approx0$ in the clean regime with $W=0$ ($L=500$), while $\Delta X_{com}/d\approx1$ for $W=1$ ($L=100$) indicates the emergence of TATP under moderate disorder strength.  For large $W$, the TATP breaks down with non-quantized COM shift ($\Delta X_{com}/d\approx0$ for large $L$). The corresponding $\Delta X_{com}$ as a function of $W$ for the lattice $L=100,500$ is shown in Fig. \ref{figS10}(c). We further perform the real-time simulation for larger lattice systems with $L=600,1000$ and obtain $\Delta X_{com}$ as a function of $W$ in Fig. \ref{figS10}(c). The result more clearly shows the moderate disorder regime for the TATP of injected light with nearly quantized COM shift. Here the deviation of the quantization is mainly due to the used single-site initial state (instead of the maximally localized Wannier state) and the finite-size effect in our simulations. \cite{Cerjan2020,Juergensen2021,QCheng2022}.

\end{document}